%% file: ms.tex
\documentclass[10pt,journal,compsoc]{IEEEtran}

\ifCLASSOPTIONcompsoc
  \usepackage[nocompress]{cite}
\else
  \usepackage{cite}
\fi

\usepackage{xspace}
\usepackage{cleveref}
\usepackage{graphicx}
\usepackage{balance}
\usepackage{fancybox}
\usepackage{dirtree}
\usepackage{multirow}
\usepackage{subcaption}
\usepackage{tabularx}
\usepackage{xcolor}
\usepackage{booktabs}
\usepackage{url}
\usepackage{soul}

\usepackage{tcolorbox} %
\tcbset{colback=gray!2}

\definecolor{darkpastelgreen}{rgb}{0.01, 0.75, 0.24}
\definecolor{darkpastelred}{rgb}{0.76, 0.23, 0.13}
\definecolor{forestgreen}{rgb}{0.13, 0.55, 0.13}
\definecolor{backcolour}{rgb}{0.94, 0.97, 1.0}
\definecolor{aliceblue}{rgb}{0.94, 0.97, 1.0}

\usepackage{listings}

\newcommand{\conclusion}[1]{%
	\begin{tcolorbox}#1\end{tcolorbox}
}

\definecolor{airforceblue}{rgb}{0.36, 0.54, 0.66}
\definecolor{bleudefrance}{rgb}{0.19, 0.55, 0.91}

\newcommand{\code}[1]{\texttt{#1}}
\newcommand{\unsafe}{\code{unsafe}\xspace}
\newcommand{\uintptr}{\code{uintptr}\xspace}
\newcommand{\cgo}{\code{cgo}\xspace}

\newcommand{\others}{et al.\xspace}
\newcommand{\snippet}[1]{\textit{``#1''}}

\newcommand{\datasetsize}{2,438\xspace}

\newcommand{\rqone}{Is \unsafe widely used in open-source Go projects?}
\newcommand{\rqtwo}{\rev{What is \unsafe used for?}}
\newcommand{\rqthree}{What are the risks of using \unsafe?}

\newcommand{\rev}[1]{#1}%

\newcommand{\revminor}[1]{#1}%

\usepackage{paralist}
\usepackage{enumerate}

\ifCLASSINFOpdf
\else
\fi

\begin{document}
\title{Breaking Type Safety in Go: An Empirical Study on the Usage of the \unsafe Package}

\author{Diego~Elias~Costa,%
        ~Suhaib~Mujahid,%
         ~Rabe~Abdalkareem,%
        ~and~Emad~Shihab~\IEEEmembership{Senior Member,~IEEE}%
\IEEEcompsocitemizethanks{\IEEEcompsocthanksitem D. E. Costa, S. Mujahid, and E. Shihab are with the Data-driven Analysis of Software (DAS) Lab at the Department of Computer Science and Software Engineering, Concordia University,
Montreal, Canada. \protect \\
E-mail: diego.costa, suhaib.mujahid, emad.shihab@concordia.ca 

\IEEEcompsocthanksitem R. Abdalkareen is with Software Analysis and Intelligence Lab (SAIL), School of
Computing, Queen's University, Canada.

E-mail: abdrabe@gmail.com

}
}

\IEEEtitleabstractindextext{%
\begin{abstract}
A decade after its first release, the Go language has become a major programming language in the development landscape. While praised for its clean syntax and C-like performance, Go also contains a strong static type-system that prevents arbitrary type casting and memory access, making the language type-safe by design. However, to give developers the possibility of implementing low-level code, Go ships with a special package called \unsafe that offers developers a way around the type safety of Go programs. The package gives greater flexibility to developers but comes at a higher risk of runtime errors, chances of non-portability, and the loss of compatibility guarantees for future versions of Go.  

In this paper, we present the first large-scale study on the usage of the \unsafe package in \datasetsize popular Go projects. Our investigation shows that \unsafe is used in 24\% of Go projects, motivated primarily by communicating with operating systems and C code, but is also commonly used as a means of performance optimization. Developers are willing to use \unsafe to break language specifications (e.g., string immutability) for better performance and 6\% of the analyzed projects that use \unsafe perform risky pointer conversions that can lead to program crashes and unexpected behavior. Furthermore, we report a series of real issues faced by projects that use \unsafe, from crashing errors and non-deterministic behavior to having their deployment restricted from certain popular environments. Our findings can be used to understand how and why developers break type safety in Go, and help motivate further tools and language development that could make the usage of \unsafe in Go even safer.

\end{abstract}

\begin{IEEEkeywords}
Go language, \unsafe, type safety, software packages, Empirical Study.
\end{IEEEkeywords}}

\maketitle

\IEEEdisplaynontitleabstractindextext

\IEEEpeerreviewmaketitle

\section{Introduction}
\input{Introduction}

\section{Background}
\label{sec:background}
\input{Background}

\section{Methodology}
\label{sec:methodology}

\input{Methodology}

\section{\rqone}
\label{sub:rq_usage}

\input{RQ_Usage}

\section{\rqtwo}
\label{sub:rq_usagepatterns}
\input{RQ_UsagePatterns}

\vspace{-0.1in}
\section{\rqthree}
\label{sub:rq_badpractices}
\input{RQ_BadPractices}

\vspace{-0.1in}
\section{Discussion}
\label{sec:discussion}
\input{Discussion}

\vspace{-0.1in}
\section{Related Work}
\label{sec:related-work}
\input{Related_work}

\section{Threats to Validity}
\label{sec:threats}
\input{Threats_to_validity}

\section{Conclusion}
\label{sec:conclusion}
\input{Conclusion}

\appendices

\ifCLASSOPTIONcompsoc
\else
  \section*{Acknowledgment}
\fi

\vspace{-0.1in}
\section*{Acknowledgment}
The authors would like to thank Matthew Dempsky, for providing invaluable feedback and for communicating our work to members of the Go Team.

\ifCLASSOPTIONcaptionsoff
  \newpage
\fi

\bibliographystyle{abbrv}
\bibliography{IEEEabrv,ms}

\vspace{-0.3in}
\begin{IEEEbiography}[{\includegraphics[width=1in,height=1.25in,clip,keepaspectratio]{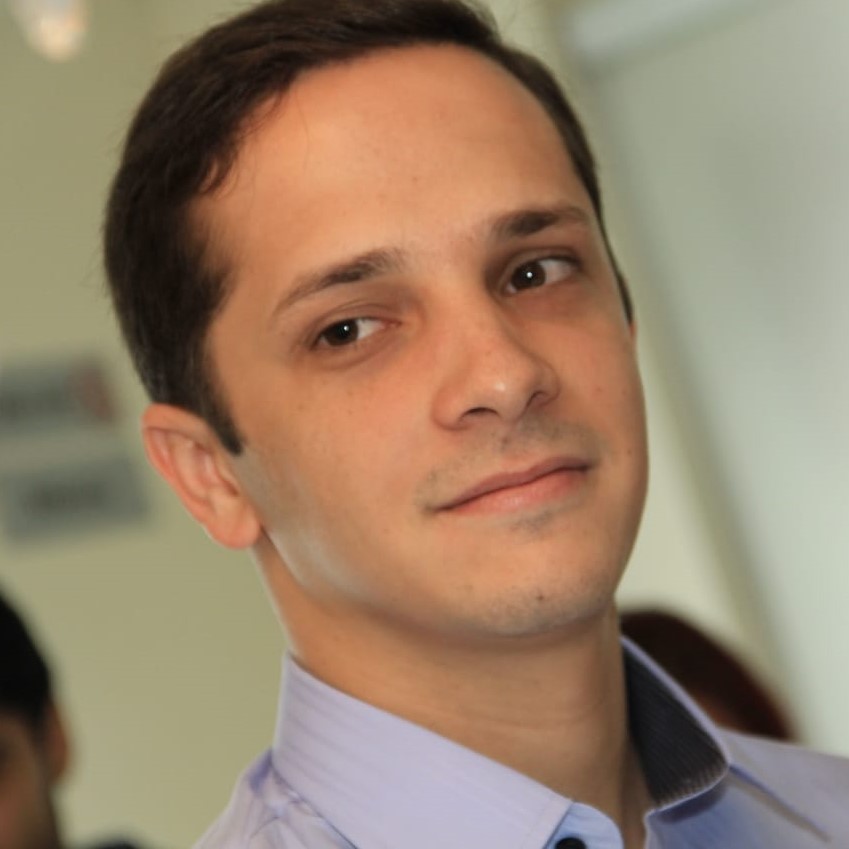}}]{Diego Elias Costa}
is a postdoctoral researcher in the Department of Computer Science and Software Engineering at Concordia University. He received his PhD in Computer Science from Heidelberg University, Germany. His research interests cover a wide range of software engineering and performance engineering related topics, including mining software repositories, software ecosystems, and performance testing. 
You can find more about him at \url{http://das.encs.concordia.ca/members/diego-costa/}.
\end{IEEEbiography}
\vspace{-0.3in}
\begin{IEEEbiography}[{\includegraphics[width=1in,height=1.25in,clip,keepaspectratio]{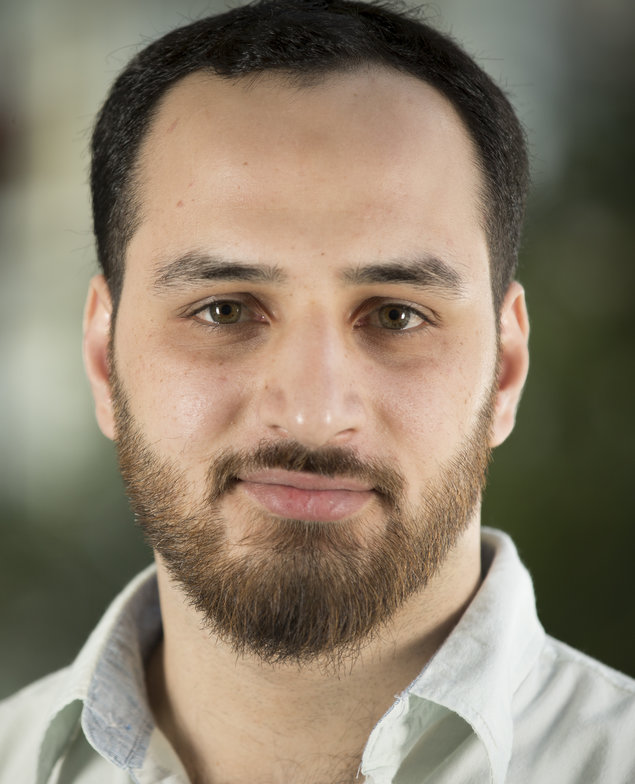}}]{Suhaib Mujahid}
is a Ph.D. student in the Department of Computer Science and Software Engineering at Concordia University. He received his master’s in Software Engineering from Concordia University (Canada) in 2017, where his work focused on detection and mitigation of permission-related issues facing wearable app developers. He did his Bachelors in Information Systems at Palestine Polytechnic University. 
His research interests include wearable applications, software quality assurance, mining software repositories and empirical software engineering. You can find more about him at \url{http://users.encs.concordia.ca/∼s mujahi}.
\end{IEEEbiography}
\vspace{-0.3in}
\begin{IEEEbiography}[{\includegraphics[width=1in,height=1.25in,clip,keepaspectratio]{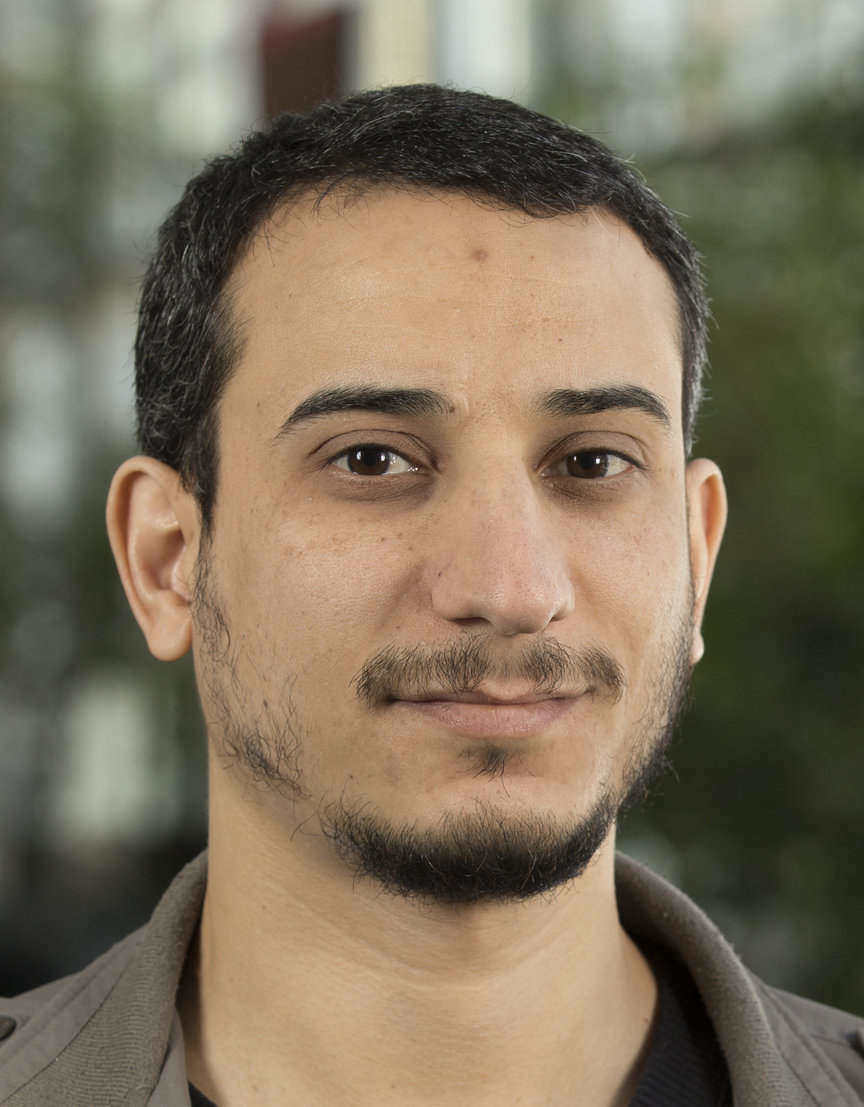}}]{Rabe Abdalkareem} received his PhD in Computer Science and Software Engineering from Concordia University. His research investigates how the adoption of crowdsourced knowledge affects software development and maintenance. Abdalkareem received his master’s in applied computer science from Concordia University. His work has been published at premier venues such as FSE, ICSME, and MobileSoft, as well as in major journals such as TSE, IEEE Software, EMSE and IST.
You can find more about him at \url{http://users.encs.concordia.ca/∼rab abdu}.

\end{IEEEbiography}
\vspace{-0.3in}
\begin{IEEEbiography}[{\includegraphics[width=1in,height=1.25in,clip,keepaspectratio]{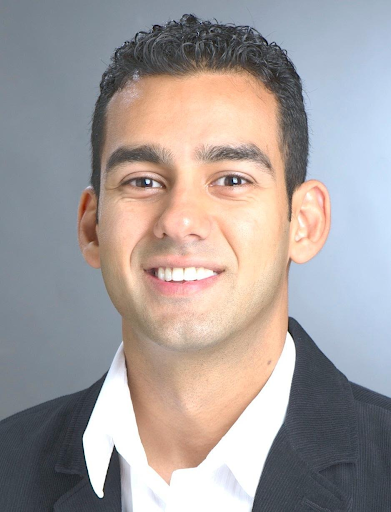}}]{Emad Shihab} is an associate professor in the Department of Computer Science and Software Engineering at Concordia University. He received his PhD from Queens University. Dr. Shihabs research interests are in Software Quality Assurance, Mining Software Repositories, Technical Debt, Mobile Applications and Software Architecture. He worked as a software research intern at Research In Motion in Waterloo, Ontario and Microsoft Research in Redmond, Washington. Dr. Shihab is a member of the IEEE and ACM. 
More information can be found at http://das.encs.concordia.ca.
\end{IEEEbiography}

\balance

\end{document}

%% file: Introduction.tex
The famous Uncle Ben's quote ``With great power comes great responsibility'' is a proverb that can be aptly applied to unsafe packages or libraries in programming languages. 
\rev{Statically-typed programming languages, such as Java, Rust, and Go restrict developers' freedom (power) by requiring every variable and function to have a type known at compile-time, in favor of a safer environment, capturing illegal type conversions and memory accesses during compilation.}
\rev{While efficient at identifying type violations, the restriction to write type-safe implementations would make it nearly impossible to write low-level code, such as system calls and pointer operations.}
Language designers address this problem by including a backdoor to violate the type system in the form of unsafe packages.
Java has \code{sun.misc.Unsafe}, Rust has \code{unsafe Rust}, and Go has the \unsafe package. 
Unsafe packages give the much needed flexibility to write type-unsafe functions, for low-level implementations and optimizations, but need to be used with extreme care by developers~\cite{Mastrangelo:2015:JavaUnsafe, Go:unsafe:online}.

Go is a statically-typed and compiled programming language released over a decade ago by engineers at Google~\cite{Go:online}. 
Its simple syntax and high efficiency has made Go one of the major programming languages in the current development landscape~\cite{Github:Octoverse, StackOverflowSurvey:2019:online}.
\rev{Go has a strong static type system, but ships with the \unsafe package~\cite{Go:unsafe:online} to offer developers the possibility of implementing low-level code.}
This package offers a way-around the \revminor{type safety} of Go %
programs, but it comes with a series of important drawbacks.
First, programs that use \unsafe may not be portable to different CPU architectures, and are not guaranteed to be compatible to future versions of Go~\cite{Go1andth70:online}.
Second, some \unsafe operations contain hidden complexities that can cause the program to exhibit non-deterministic errors.
For instance, Go contains pointers and a Garbage Collector (GC), hence, manipulating (unsafe) pointers without proper care may cause the GC to release unwanted variables~\cite{Go:unsafeptr:online}.

Given that the benefits of writing type-safe code are well-known and lifting this safety net puts programs at a higher risk of runtime errors~\cite{Mastrangelo:2015:JavaUnsafe},  why do developers break \revminor{type safety} in Go?
There is no shortage of articles in the web warning against the perils of using \unsafe~\cite{GoWhatis70:online, Go:unsafe:online, Go:bytes2string:online}, and maintainers of Go have had extensive debates over the need and consequences of keeping this package in the language~\cite{Guarante80:online}.
However, it is hard to derive effective measures on how to handle the risks and benefits of \unsafe package without knowing the extent in which developers use it, what they use it for, and what are the real risks of breaking \revminor{type safety} in Go projects.
Our study is a step towards acquiring this understanding.

In this paper, we perform a mix-method empirical study involving \datasetsize popular Go open-source projects. We first develop a parser to identify usages of the \unsafe package throughout the development history of Go projects. \revminor{Then, we perform a manual analysis to evaluate the prevalence of \unsafe usage and their consequences in the most popular Go projects. }
Our study focuses on answering the following research questions:

\setdefaultleftmargin{25pt}{}{}{}{}{}
\begin{enumerate}
	\item[\textbf{RQ1:}] \textbf{\revminor{(Prevalence of \unsafe)} \rqone} We found that 24\% of the studied Go projects use \unsafe at least once in their code-base.
	The usage of \unsafe grows with the project size, with projects using \unsafe on 17\% of their packages. 
	The package \unsafe is used in a wide variety of project domains (e.g.,  networking, blockchains, containers, development tools, databases). 
	\rev{We also found that projects that rely \revminor{most} heavily on the \unsafe package, with more than 100 call-sites, are the ones that implement bindings to other platforms and/or programming languages.} %

	\item[\textbf{RQ2:}] \textbf{\revminor{(Prevalence of \unsafe)} \rqtwo} 
	We catalogued six groups of usage-patterns related to \unsafe in Go. 
	The majority of \unsafe usages were motivated by integration with operating systems and C code (45.7\%), but developers also frequently use \unsafe to write more efficient Go code (23.6\%). 
	Less frequently, developers use \unsafe to perform atomic operations, dereferrence values through reflection, manipulate  memory addresses and get the size of objects in memory.

	\item[\textbf{RQ3:}]\textbf{\revminor{(Consequences of \unsafe)} \rqthree} 
	\rev{At least 6.6\% of the investigated projects that use \unsafe have invalid pointer conversions, a risky usage of the API that may cause crashing bugs and lead programs to unexpected runtime behavior.}
	Developers report a variety of issues caused by using \unsafe, from having their deployment restricted by environment platforms, to crashing and non-deterministic errors found in production.

\end{enumerate}

Our study provides empirical evidence that contributes towards a safer Go programming language. 
Our findings show that the usage of \unsafe is widespread, covering all sorts of project domains, and is motivated by integration with Operating Systems, C programs, and more efficient implementations.
The usage patterns we catalogued can be used by tool designers to better assist developers when performing unsafe operations, as well as guiding standard packages to improve documentation to cover the most common use-cases.
Furthermore, our results indicate that even popular projects are not immune to well-known pitfalls associated with the package usage and that projects that use \unsafe report a variety of unsafe-related issues.
Our results also help developers at identifying potential risks and pitfalls to avoid when using the \unsafe package.

This paper is organized as follows: 
\Cref{sec:background} introduces the \unsafe package and the concepts that we use throughout the paper.
\Cref{sec:methodology} presents our methodology used for our study.
The results of our study are presented in three sections, %
\Cref{sub:rq_usage}, \Cref{sub:rq_usagepatterns}, and \Cref{sub:rq_badpractices} presents the results of RQ1, RQ2, and RQ3, respectively.
We discuss our findings and implications in \Cref{sec:discussion}.
Then, in \Cref{sec:related-work} we present the related work and discuss the threats to the validity of our results in \Cref{sec:threats}.
Finally, we conclude our study in \Cref{sec:conclusion}.

%% file: Background.tex
\begin{figure*}[tbh]
	\centering
		\includegraphics[width=\linewidth, trim=.5cm 8.4cm .5cm  1.8cm, clip]{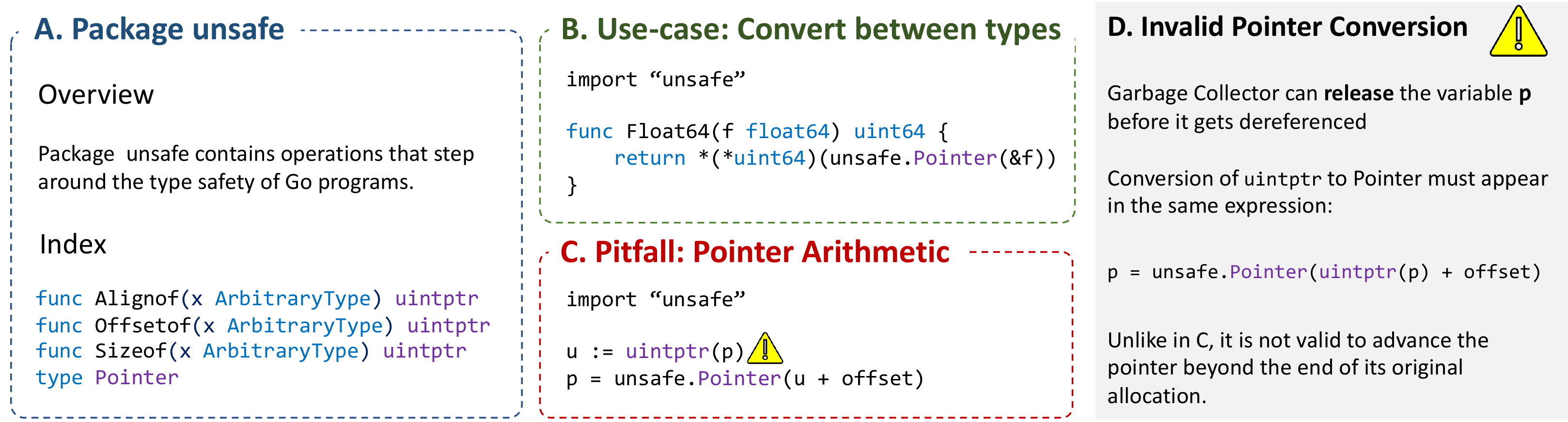}
	\caption{The \unsafe package API (A), an example of how to use \unsafe to convert between types in Go (B) and a pitfall related to performing pointer arithmetic in Go (C).
	On the right we explain the invalid pointer conversion issue and give an example of solution for a safe pointer arithmetic. All examples were taken from the official documentation~\cite{Go:unsafe:online}}
	\label{fig:unsafe-background}
\end{figure*}

In this section, we introduce the \unsafe package API and illustrate some of its use-cases. We describe the risks of using \unsafe and dive into a unsafe-related pitfall that may cause programs to crash or exhibit a non-deterministic behavior. 

\subsection{Type safety in Go and the \unsafe package}

Go is a statically typed language. That is, the type of variables in a Go program are known at compile time, enabling  the compiler to verify type inconsistencies and incompatible type conversions.
A compiled Go program is guaranteed to be type-safe which prevents a myriad of runtime issues, unless developers use the \unsafe package. 
As described in the official documentation shown in \Cref{fig:unsafe-background}, the \unsafe package offers a step around the type safety of Go programs~\cite{Go:unsafe:online}. 
The \unsafe package is quite compact, containing only three functions and one Pointer type in its API, shown in the Index of \Cref{fig:unsafe-background}-A.
By using the \unsafe package, developers have the flexibility needed to develop low-level code, such as full control over pointers (C-style), the ability to read and write arbitrary memory, access to object sizes, and the possibility to convert between any two types with compatible memory layouts (shown in \Cref{fig:unsafe-background}-B).
\rev{Compared to unsafe libraries from other programming languages like the \texttt{sun.misc.Unsafe} from Java and \texttt{unsafe Rust}, the Go \unsafe package is still much more restricted. By using their respective unsafe APIs, in addition to breaking \revminor{type safety}, developers can violate method contracts in Java~\cite{Mastrangelo:2015:JavaUnsafe} or use inline assembly functions in Rust~\cite{Evans:2020:Rust}.}

\subsection{Risks of using the \unsafe package}
\label{sub:backgrond-risk-unsafe}

In practical terms, breaking \revminor{type safety} means lifting the safety net that a compiler provides for developers in exchange for full-control of reading and writing memory. 
This by itself, puts developers at a higher risk of making mistakes that will ripple through to the production environment.
\rev{For instance, converting between incompatible types may cause the program to eventually crash or misinterpret the memory layout of a variable causing unexpected program behavior.}
Aside from this, the \unsafe package comes with a series of drawbacks described in its documentation~\cite{Go:unsafe:online}, that need to be taken into careful consideration by developers:

\noindent
\textbf{Non-Portability:}
A regular package in Go can be compiled to different CPU architectures and operating systems without any changes in the code. 
Projects that use \unsafe, however, may need to account for differences in the CPU architectures and it is up to developers to keep their packages portable.
For instance, by traversing an array of integers using \unsafe, developers need to account for the integer size which differs in x86 and x64 architectures.
\rev{Another example is the reliance on system calls, which are accomplished by using both \texttt{syscall} with \unsafe packages and are not often portable to different operating systems.}

\noindent
\textbf{Not Protected by the Compatibility Guideline:}
Programs that are written in Go have the guarantee to work in future versions of the programing language, as long as they follow the compatibility guideline established in Go 1~\cite{Go1andth70:online}. 
However, using the \unsafe package breaks this compatibility guarantee, as \unsafe exposes the internal implementations of the language, which may change in future versions.
Hence, programs that rely on \unsafe are not guaranteed to work in future implementations of Go.

\subsection{Invalid Pointer Conversion}
\label{sub:unsafe-misuse}

The \unsafe package provides an extensive documentation on how to properly use the package and some of the pitfalls developers must avoid when breaking \revminor{type safety}~\cite{Go:unsafe:online}.
A pitfall that is well-described in the package documentation is the invalid pointer conversion.
As a rule of thumb, converting pointer addresses back to pointer variables is not valid in Go language.
Pointer memory addresses in Go are of the type \uintptr (unsigned integer pointer), which is a simple integer type, holding no reference or pointer-semantics. 
Since Go has a Garbage Collector that releases memory that is not referenced by any variable in the program, a variable \uintptr holding the address of a pointer will not prevent the garbage collector \revminor{from releasing} the said pointer.
Consequently, there is no guarantee that a memory address (type \uintptr) contains a valid Go pointer variable to be dereferenced~\cite{Go:unsafeptr:online}.

In some cases, however, there is a valid need to manipulate memory addresses and dereference them back to pointers. 
For instance, to perform pointer arithmetic operations (C-style) to traverse an array as shown in \Cref{fig:unsafe-background}-C.
Developers need to be aware of the intricacies of manipulating low-level pointers in a language that contains a Garbage Collector, to prevent the Garbage Collector from releasing their variables in the middle of their operations.
We show in \Cref{fig:unsafe-background}-D an example on how to properly handle pointer arithmetics in Go.
In this case, developers should never store their \uintptr into an intermediate variable (\code{u}), because at this point in the execution of a program the variable \code{p} can be released by the Garbage Collector. 

This example shows that the pitfalls of handling \unsafe are not always intuitive.  %
To make matters worse, the issues that arise from using the invalid pointer conversion are non-deterministic and are unlikely to be encountered during software testing when memory pressure tends to be small.

%% file: Methodology.tex
To understand the prevalence of \unsafe and its impact in open source Go projects, our study has three main goals:

\setdefaultleftmargin{12.5pt}{}{}{}{}{}
\begin{enumerate}
	\item Understand the extent in which projects use \unsafe package in their source-code~(\Cref{sub:rq_usage}).  To achieve this, we identify projects that use \unsafe with a parser,  investigate whether the usage of the package changes as the project evolves and what are the categories of projects that more frequently rely on \unsafe.
	
	\item Understand what developers use \unsafe for (\Cref{sub:rq_usagepatterns}). We investigate what are the most used features from the \unsafe API and manually extract the most common usage patterns of \unsafe in Go projects.
	
	\item Understand the risks that using \unsafe entails to projects (\Cref{sub:rq_badpractices}).
	We examine \revminor{whether projects have invalid pointer conversions} in their code and qualitatively evaluate issues related to the \unsafe package. %
	
\end{enumerate}

\begin{figure*}[tb]
	\centering
	\includegraphics[width=\linewidth, trim=2cm 8.5cm 2cm  1cm, clip]{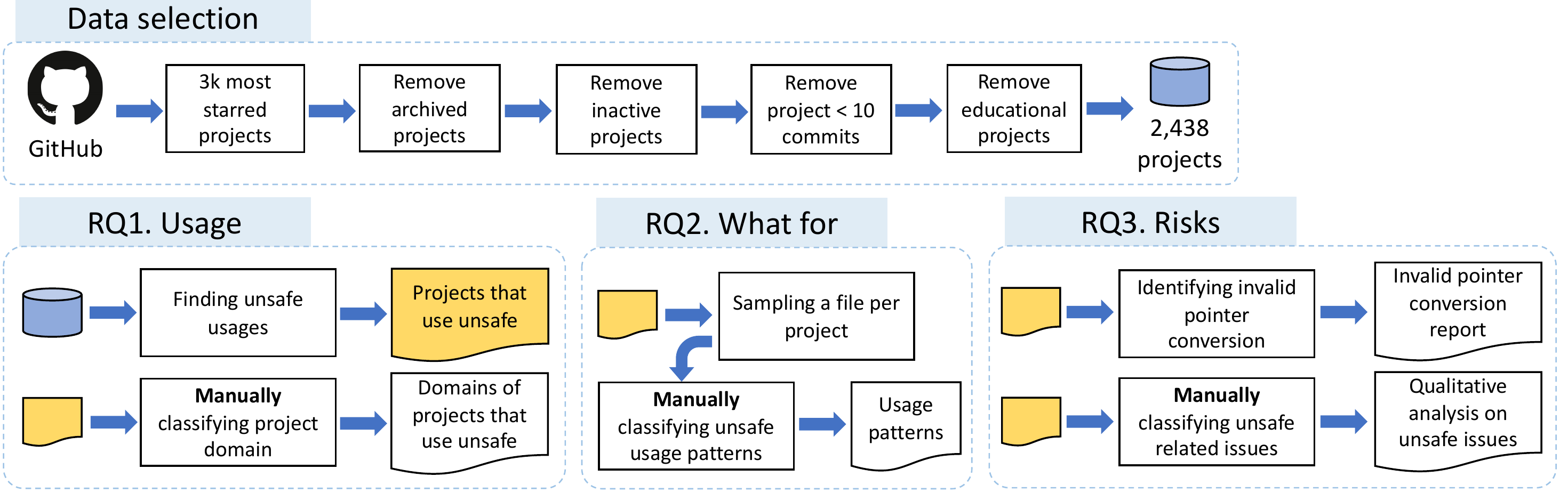}                    
	\vspace{-0.21in}   
	\caption{Overview of the methodology adopted in our study.}
	\label{fig:methodology}             
\end{figure*}

In the following subsections, we describe the methodology used to achieve the goals of our study.
We present the overview of our methodology in \Cref{fig:methodology}.

\subsection{Study dataset}

To investigate the usage of \unsafe on a large-set of popular Go projects, 
we first started by querying the GitHub REST API\footnote{https://developer.github.com/v3/} to identify a representative set of open source Go project. To do so, we selected a set of the top 3,000 most starred Go repositories,
as the number of stars is an indicator of the project popularity within GitHub repositories~\cite{Borges:GitHubPopularity, Borges:GitHubPopularityFactors}.
Our dataset was collected on October 2nd, 2019.

Even within highly starred repositories, we may find inactive repositories or repositories not related to software development.
To get a representative set of high-quality and active Go software development projects, we follow the methodology recommended by  previous work~\cite{Kalliamvakou:2016:GitHubPerils} to further curate our dataset through the following criteria: 
\begin{enumerate}
	\item We filter out 89 archived projects, as these projects are no longer maintained by their developers. 
	Archived projects are identified via a flag in the project's metadata.
	\item We removed 371 inactive projects, by filtering out projects that have had no contribution 12 months prior to the data collection date (after October 2018). 
	\item We removed 22 projects with less than 10 commits in total, as these projects tend to be too young and are not representative of the projects we aim to investigate.
	\item We also removed 56 educational repositories, which are fairly common on popular projects datasets. These projects are described as published books, programming courses or any learning material, which are invaluable to the community, but are not representative of typical Go software. We removed these by manually inspecting project descriptions of our entire dataset.
	\item Furthermore, 24 projects could not be cloned automatically by our scripts (e.g., invalid URL, project no longer available) and were removed from our study.
\end{enumerate}

This process yields a dataset composed of \datasetsize Go projects. \Cref{tab:stats-dataset} shows the summary statistic of the selected Go projects in our dataset. 
As shown in \Cref{tab:stats-dataset}, our dataset contains very popular projects (median of 857 stars and 110 forks), with a relatively long development history (median of 3.5 years of development).
\rev{The dataset contains eight projects with less than 150 LOC (minimum of 77 LOC), but in its large majority are composed of medium to large software projects (median of 9,6 KLOC).
We do not focus on filtering projects based on their size, as we think the perils of breaking type safety affects projects of all sizes.}

\begin{table}[t]
	\centering
	\caption{Statistics of the projects in our dataset.}
	\vspace{-0.05in}
	\label{tab:stats-dataset}
	\input{tables/stats-dataset}
\end{table}

\vspace{-0.1in}
\subsection{Identifying \unsafe usages}

To find whether projects make use of \unsafe in their source-code, we build a parser for Go source code, using the support of Go's native \code{ast} package\footnote{https://golang.org/pkg/go/ast/}. 
Our parser first analyzes the source code of each Go file (files with .go extension) in the selected projects and build an abstract syntax tree (AST) for each file. 
Then, the parser inspects the AST of Go files and identifies the function calls and type references to \unsafe.
First we run our parser at the latest snapshot of the projects obtained during our data collection.
In addition, we also want to understand how the usage of \unsafe changes as the project evolves.
To that aim, we use our parser to analyze the snapshot of the first commit of each month in the projects' history, and  
For this analysis, we only consider the default branch of each project, identified via the GitHub API.

\subsection{Classifying projects' domain}

To complement the analysis on the usage of \unsafe, we investigate the domain of the projects that use the \unsafe package, as the domain has a direct influence on the need for breaking \revminor{type safety}.
Intuitively, we expect projects that demand low-level implementation and optimizations, such as databases and file systems, to depend more on \unsafe implementations than other type of projects such as data structure libraries and web applications.
To identify the projects' domain, we manually inspect the description and documentation in their GitHub page and classify each project into a dominant domain, e.g., database, compiler, web server, development tool.
The three-first authors classify the repositories using an open card-sort method~\cite{Fincher_2005OpenCardSort}, where labels are created during the labeling process and each new label is discussed among annotators and retroactively applied to previously classified projects.
When different labels were assigned to the same project we \revminor{discussed them} to reach a consensus.

\subsection{Classifying \unsafe usage patterns}
\label{sub:classifying-patterns}

To investigate what developers aim to accomplish using \unsafe in RQ2, we need to understand the most frequent usage patterns associated with the \unsafe package.
Each usage pattern may offer a rationale that will help us understand the proximate reason as to why developers opted for breaking \revminor{type safety} in their projects.
To extract high-level usage patterns from the projects' source code, we resort to in-depth manual analysis of the code, documentation, and commit messages.
It is important to notice that this analysis is very time consuming, as annotators need to recognize the context in which \unsafe is being used, search for clues that indicate the reason behind the \unsafe in commit messages and code documentation.
Hence, we decide to perform such analysis on a statistically significant random sample of the projects that use \unsafe.
This sample is drafted to provide us with a representative set of the projects that rely on unsafe with 5\% confidence interval at 95\% confidence level.

We expect some projects to contain hundreds of \unsafe call-sites while most projects may use \unsafe sporadically in their code, given the risks associated with the package.
We want to identify patterns across projects and avoid biasing our analysis towards projects that rely more extensively on \unsafe in their implementation.
Hence, we perform a second sampling by randomly selecting a single file from each project to our analysis.

It is important to highlight that we opt to conduct this particular analysis at the file level as opposed to package-level for the following reasons:
1) Files are more fine-grained than packages and are expected to have a more cohesive structure where we could more easily derive the usage pattern; 
2) We can analyze the context of usage of a single file, with support of \texttt{gitblame} to futher inspect commit messages, without the need \revminor{to inspect} method calls across different files, which would impose a prohibitive time cost to this analysis.

The first two authors independently labeled each file (one per project) using an open card-sort method~\cite{Fincher_2005OpenCardSort}.
Hence, similarly to the analysis of projects' domains, labels are created and assigned while inspecting the usage in source-code and git commits, and every new label is discussed among annotators and if necessary, retroactively applied to previously labelled usages.
We assess the agreement of both annotators using the Cohen-Kappa metric~\cite{Mary:2012:kappa} and both annotators discussed and merged the results.

\vspace{-0.1in}
\subsection{Identifying invalid pointer conversions}

\revminor{In RQ3, we investigate the risks to projects caused by using the \unsafe package.}
As discussed in \Cref{sec:background}, a major pitfall of using unsafe pointers is being unaware of issues related to invalid pointer conversions. 
This issue can lead to non-deterministic errors in programs, as the Garbage Collector may release unintended variables and causing the program to crash.
We want to investigate whether invalid pointer conversions do occur in the most popular Go projects. 
To automatically identify suspicious cases of invalid pointer conversions, we resort to existing static analyzers, as this issue is well-documented and covered by a go code analyser: Vet~\cite{vetTheGo72:online}.
Go Vet was created by maintainers of the Go language and is shipped natively with Golang. 
\rev{It identifies suspicious cases of invalid pointer conversion when run with the flag \texttt{unsafeptr}. 
Go Vet parses the code and applies a set of heuristics based on known ill-formed expressions (such as the one in \Cref{fig:unsafe-background}), exporting every suspicious case as a json file.
As with any heuristic-based tool, Go Vet may miss cases that do not fall into the most known patterns of invalid pointer conversions.}

\vspace{-0.1in}
\subsection{Classifying \unsafe related issues}

Another method for understanding the risks associated with \unsafe usage is to investigate the issues open in Github related with the package use.
To analyze and classify the issues related to \unsafe, we first mine all issues from the repository of all projects that use \unsafe. 
In GitHub, ``issue" is an umbrella term that encloses pull requests, bugs, questions to maintainers and requests for new features. 
\rev{We find candidate issues for our analysis by applying a keyword search for ``unsafe" and two variations (``un-safe'' and ``not safe") in the issue title. }
The keyword approach is prone to false-positives, especially given that the key-words are used in several different contexts (e.g., multi-threading), hence, we need to filter out false-positives from this candidate set.
The first author manually inspected each issue and removed the false-positives. 
Then the first two authors proceed to analyze the issue title, body and related commit code to group issues based on their similarities. 
Similarly to the previously described methods, we resort to the open-card methodology~\cite{Fincher_2005OpenCardSort} and evaluate the interrated agreement using the Cohen-Kappa interrater method~\cite{Mary:2012:kappa}.
We discuss the disagreement in a second round to reach a consensus in the classification.
Note that the goal of this analysis is not to find all issues related to \unsafe, but  rather to classify a sample of possible unsafe-related issues to provide qualitative insights about the problem related to use the \unsafe package.

\vspace{-0.1in}
\subsection{Replication Package}

To facilitate verification and advancement of research in the field, we provide a replication package containing the list of projects analyzed, the data extracted from the projects, and the scripts used to process and analyze our RQs~\footnote{\url{https://zenodo.org/record/3871931}}.

%% file: tables/stats-dataset.tex
\begin{tabular}{lrrrr}
	\toprule
	\textbf{Statistics} &       \textbf{Mean} &    \textbf{Min} &      \textbf{Median} &           \textbf{Max} \\
	\midrule
	Age (months) &      44.17 &   1 &    43 &        119 \\
	Stars        &   1,967.81 & 314 &   858 &     64,079 \\
		Forks        &     276.71 &   2 &   109 &     20,437 \\
	LOC          & 210,217.09 &  77 & 9,606 & 16,579,983 \\
	Commits      &   1,028.48 &  11 &   261 &     97,504 \\
	\bottomrule
\end{tabular}

%% file: RQ_Usage.tex
\noindent\textbf{Motivation:} 
\rev{We start the study by investigating to what extent developers use \unsafe in open-source Go projects. 
This will help us understand how frequently developers abandon type safety guarantees to implement their programs; \revminor{how \unsafe usages change as the project evolves and which domain of projects make more use of \unsafe}.
This understanding is crucial to motivate the development of better solutions related to \unsafe usage in Go.
We analyze this question under three complementary aspects:
}

\setdefaultleftmargin{12.5pt}{}{}{}{}{}
\begin{enumerate}
	\item \textbf{Usage:} How often does a Go project uses \unsafe in their source-code?
	This will help us understand how frequently developers abandon type safety guarantees to implement their programs.
	
	\item \textbf{Trend:} Does the usage of \unsafe change over the evolution of a project? 
	This analysis will give us insights on whether \unsafe usage increases and spreads to multiple packages as the project evolves, or if developers make the conscious effort of isolating \unsafe to mitigate its risks.
		
	\item \textbf{Domain:} What domain of projects rely on \unsafe?
	With this analysis, we aim at identifying what categories of projects are more susceptible to breaking type safety.

\end{enumerate}

\vspace{-0.1in}
\subsection{How often does a Go project use \unsafe?}

\noindent\textbf{Approach:}
We run our parser to identify \unsafe usages in every Go file present at the latest snapshot of projects in our dataset, but filter out usages identified in the source-code of the project dependencies. 
\revminor{A commonly employed practice in Go projects for managing dependencies is to include all dependencies in the projects structure, in a folder called ``vendor''. }
Hence, we exclude all reports of \unsafe usage originating from vendor folders.

\begin{table}
	\centering
	\caption{Statistics on the projects that use \unsafe. }
	\vspace{-0.1in}
	\label{tab:rq1:descriptive}
	\include{tables/rq1-descriptive}
	\vspace{-0.1in}
\end{table}

\begin{figure*}
	\begin{subfigure}{.33\linewidth}	
		\centering
		\includegraphics[trim=0 0 1.4cm  0, clip,width=\linewidth]{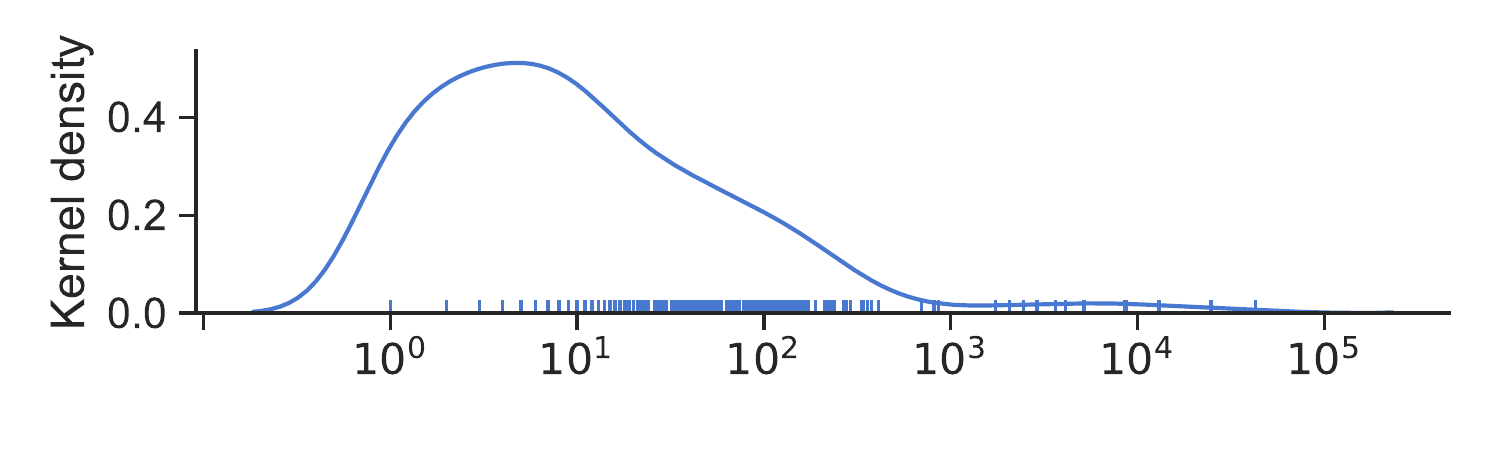}
		\caption{\# of \unsafe call-sites.}
		\label{fig:rq1-distribution-callsites}
	\end{subfigure}
	\begin{subfigure}{.33\linewidth}	
		\centering
		\includegraphics[trim=0 0 1.4cm  0, clip,width=\linewidth]{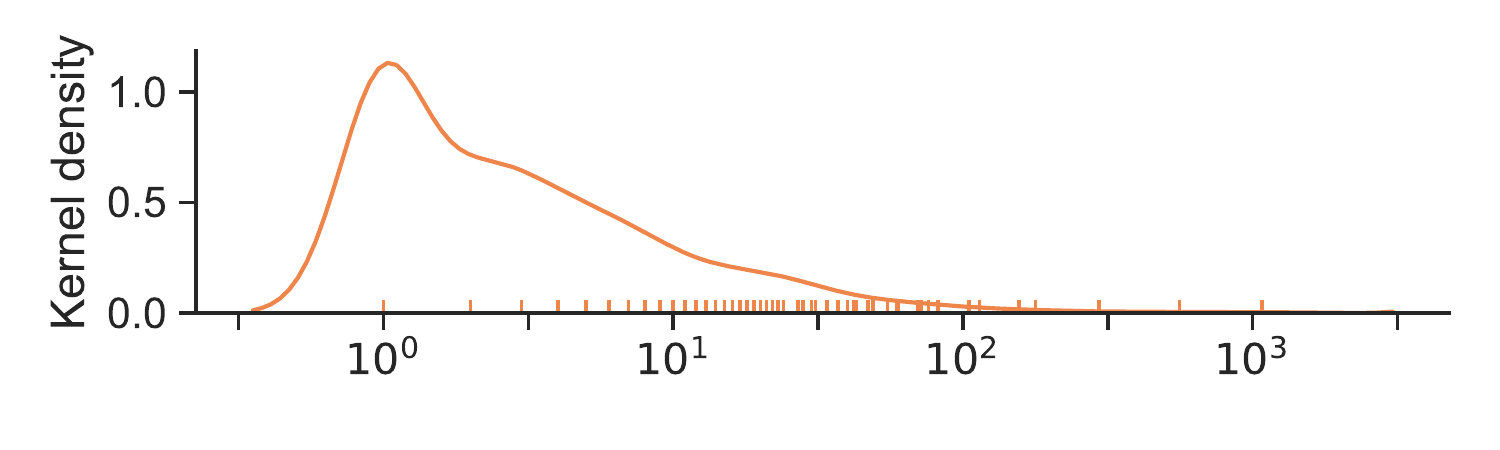}
		\caption{\# of files with \unsafe call-sites.}
		\label{fig:rq1-distribution-files}
	\end{subfigure}
	\begin{subfigure}{.33\linewidth}
		\centering
		\includegraphics[trim=0 0 1.4cm  0, clip,width=\linewidth]{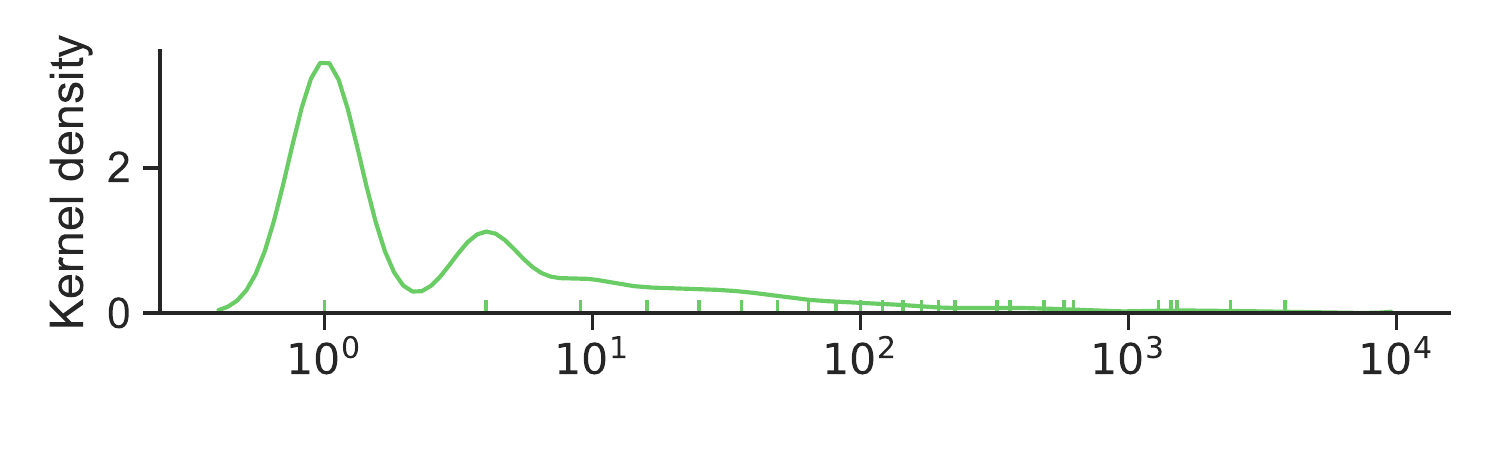}
		\caption{\# of packages that depend on \unsafe.}
		\label{fig:rq1-distribution-packages}
	\end{subfigure}
	
	\caption{Distribution of \unsafe usage in three granularity levels (call-site, files and packages) per project as a KDE density estimation plot. Note that the x-axis is logarithmic.}
	\label{fig:rq1-distribution-projects}
	\vspace{-0.1in}
\end{figure*}

\noindent\textbf{Results:}
\Cref{tab:rq1:descriptive} shows that from the \datasetsize evaluated Go projects, \textbf{592 (24\%) make use of \unsafe directly in the project source-code}. 
The extent in which projects use \unsafe varies considerably.
In \Cref{fig:rq1-distribution-projects} we show the distribution of the \unsafe usage in our dataset as a Kernel-density estimation plot, under three granularities: call-sites, files, and packages. 
The call-sites plot accounts for every \unsafe operation called in a project, the files indicate how many files depend on \unsafe and the package plot shows how many modules in Go depend on \unsafe. 
As evidenced by the \Cref{fig:rq1-distribution-callsites} peak, the majority of projects (57\%) contains \revminor{fewer than 10 calls} to \unsafe operations in their source-code.
Consequently, most projects concentrate their \unsafe calls in at most 2 files and a single package, keeping the \unsafe usage well-localized in the code (see the distribution shown in Figures \ref{fig:rq1-distribution-files} and \ref{fig:rq1-distribution-packages}).
Yet, we found that 69 projects (2.8\%) in our dataset rely heavily on \unsafe in their project, with more than 100 call-sites present in their source-code.

\vspace{-0.1in}
\subsection{Does the usage of \unsafe change as projects evolve?}

\noindent
\textbf{Approach:}
Since we want to analyze whether and how the usage of \unsafe changed over the course of the project development, we examine the use of \unsafe package at different snapshot of the projects history.
As our dataset contains in median projects with almost 4 years of development (see \Cref{tab:stats-dataset}), we focus on analyzing the trend of \unsafe usage on the period between January 2015 to September 2019 (a month before our data collection time).
In addition, to perform a sound analysis of the usage of \unsafe over the years, we only conduct this analysis on projects that fulfill the following criteria: 1) projects that have at least 10 \unsafe operation calls at \revminor{the latest snapshot of their source code}, as these will show a more meaningful evolution of \unsafe usage; and \rev{2) projects that were being actively developed during this entire period, i.e., projects with at least one commit pushed in each year. It is important to notice that we applied these criteria to avoid skewing our results towards a particular time frame.}
For instance, younger projects could skew our analysis as a higher number of projects would be accounted for in the most recent years.

\begin{figure}
	
	\begin{subfigure}{\linewidth}
		\centering
		\includegraphics[width=.75\linewidth]{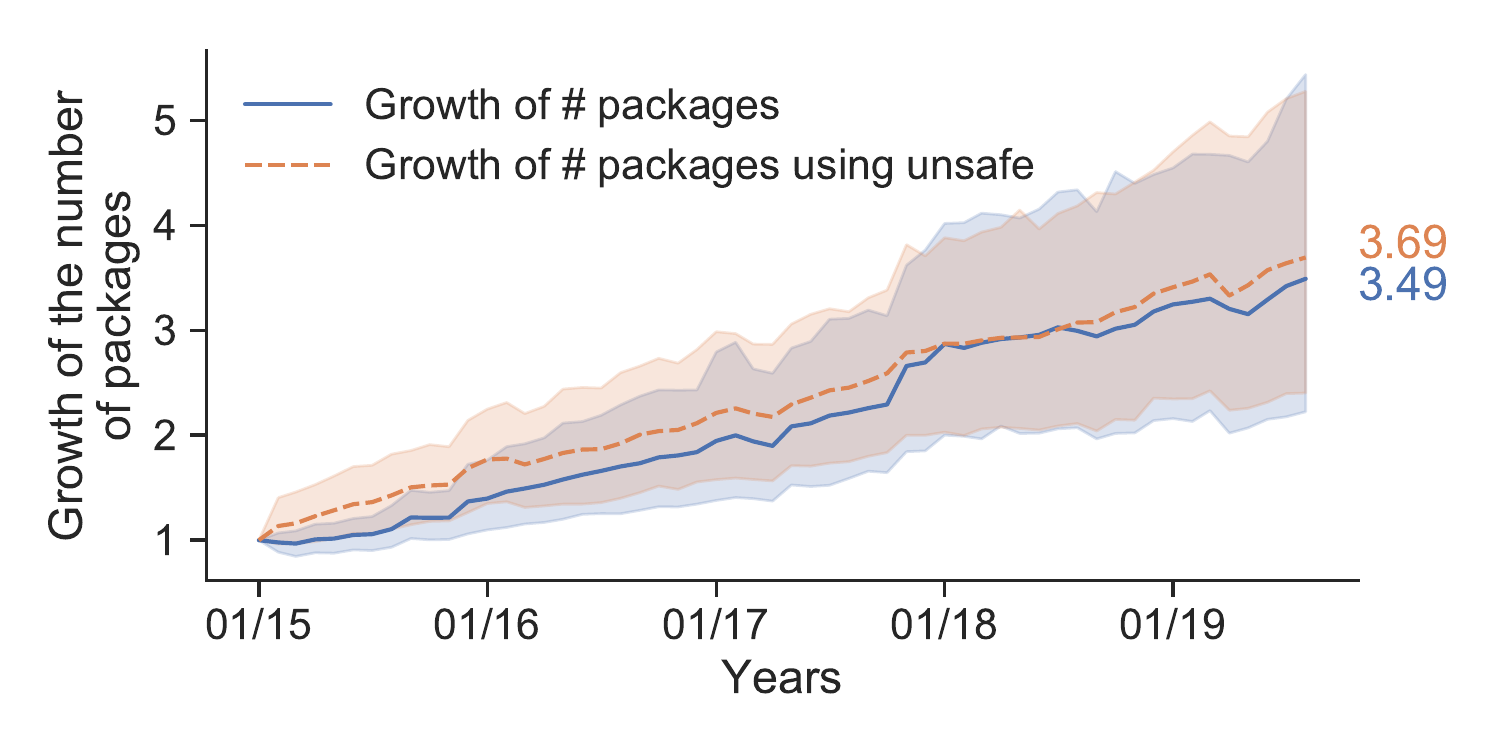}
		\vspace{-0.1in}
		\caption{Evolution of 270 projects compared to their state in January 2015 (baseline) in terms of the number of packages and number of packages using \unsafe.}
		\label{fig:growth-packages}
	\end{subfigure}

	\begin{subfigure}{\linewidth}
		\centering
		\includegraphics[width=.75\linewidth]{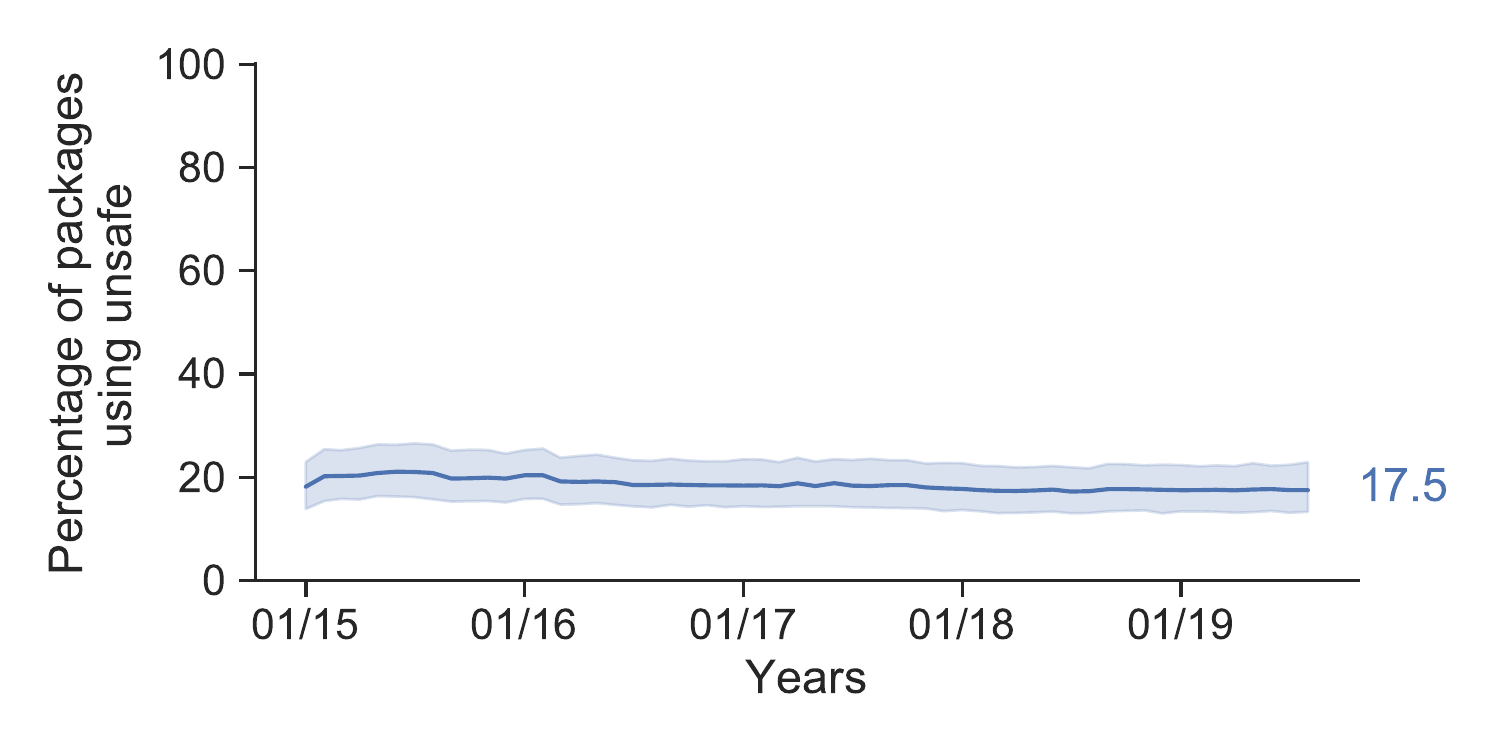}
		\vspace{-0.1in}
		\caption{Evolution of the percentage of packages that use \unsafe in 270 projects.}
		\label{fig:growth-unsafe}
	\end{subfigure}
	\caption{Analysis of projects' evolution in terms of number of packages and the share of packages that rely on \unsafe. }
	\label{fig:rq1-trend-all}
\end{figure}

\noindent
\textbf{Results:}
\rev{Figure~\ref{fig:rq1-trend-all} shows the evolution of the 270 projects in terms of number of packages and packages using \unsafe (\Cref{fig:growth-packages}) together with the percentage of packages that rely on \unsafe (\Cref{fig:growth-unsafe}). 
In both plots, the thick line represents the mean value and the colored area shows the 95\% confidence interval of the data at each month.}

\rev{As shown in \Cref{fig:growth-packages}, the number of packages using \unsafe grows similarly to the growth of total number of packages in a project. 
As the number of packages increased by 3.5 times in our studied projects, the number of packages using unsafe also increased at a similar rate (3.7 times).
This indicates that the usage of \unsafe evolves linearly as the project size increases, in the evaluated projects.
In addition, \Cref{fig:growth-unsafe} shows the proportion of packages using \unsafe, which remains steadily around 17\% as projects evolve.
Our results paint a \revminor{mixed} picture; while the dependency of \unsafe does not spread out to a larger proportion of packages, it has also not decreased as projects grow in average 3.5 times, since January 2015.}

\subsection{What domain of projects rely the most on \unsafe?}

\noindent
\textbf{Approach:}
We want to understand whether there is a difference in the domain of projects that use \unsafe sporadically, against projects that rely more heavily on breaking type-safety.
Hence, after categorizing the projects' domain, we group the projects that use \unsafe into two groups: 
in the first group we include all projects that use \unsafe, and in the second group we focus on projects that rely heavily on \unsafe, i.e., the 69 projects containing more than 100 \unsafe call-sites.

\begin{figure}
	\includegraphics[width=\linewidth]{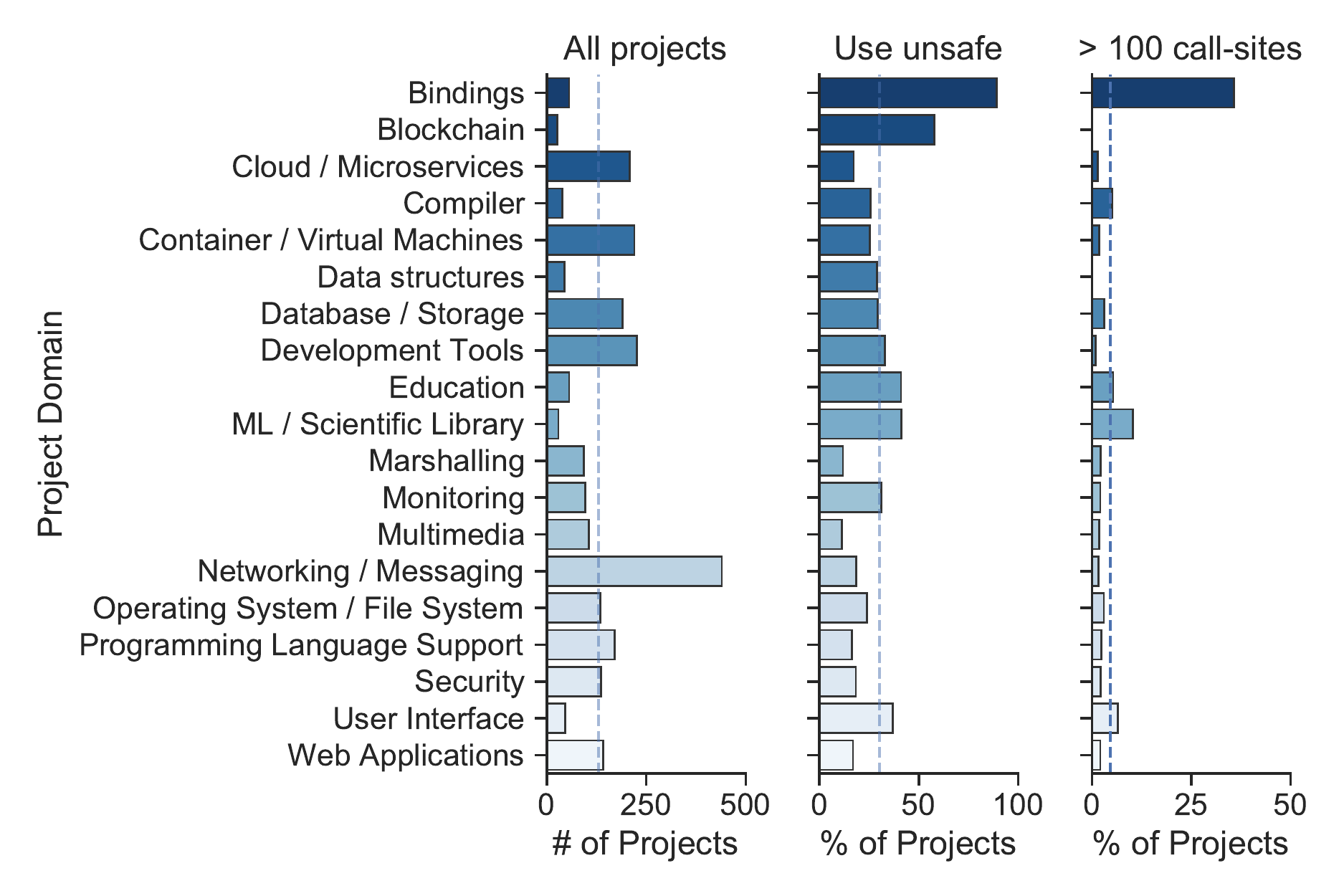}
	\caption{Distribution of the number of projects per domain (left), accounting for projects that depend on \unsafe (middle), and projects that rely heavily on \unsafe. \revminor{We highlight the averages across all project domains as dashed lines}. 
	}
	\label{fig:rq1-project-categories}
	
\end{figure}

\noindent
\textbf{Results:}
\rev{\Cref{fig:rq1-project-categories} shows the distribution of project domains of all projects in our dataset (left), percentage of projects per domain that use \unsafe (middle) and the percentage of projects that contain more than 100 \unsafe call-sites (right).
First, we observe that our dataset contains a predominance of Networking/Messaging projects in our dataset (440 projects), followed by Development Tools (227 projects) and Container/Virtual Machines (220 projects).
As our results show, \textbf{projects from 20 different domains make use of \unsafe}.
The domains with the highest percentage of projects using \unsafe are Bindings (89\%) and Blockchain (57\%) projects, followed by ML/Scientific Library (41\%). 
Several other domains that are more directly related to providing tools for software development, such as Development Tools (33\%), Monitoring (31\%), Database/Storage (29\%), and Data Structures (29\%), have more than a quarter of their projects relying on \unsafe to some extent. }

\rev{Most notably, \textbf{projects that rely heavily on \unsafe (more than 100 call-sites) represent 35\% of all Binding projects in our dataset, the highest share in any category.}
Binding projects are projects that aim at bridging the Go language to libraries and platforms written in different programming languages.
These projects, such as Gotk3~\cite{gotk3:online} a bindings for the Graphical Interface framework GTK~\cite{gtk:online}, integrate with platforms not written in Go and use \unsafe to implement functions that communicate with operating systems and C code.
Aside from Bindings projects, other domains have a handful of projects with more than 100 \unsafe call-sites, such as Networking (7 projects) and Database (6 projects) domains. 
Example of Networking and Database projects that rely heavily on \unsafe are the Networking and Security service Cilium~\cite{Cilium:online} and the cloud-native SQL database CockroachDB~\cite{cockroachdb:online}.
Interestingly, only two domains do not have projects with more than 100 \unsafe call-sites, which are Data structures and Blockchain.}

\conclusion{
\textbf{Summary of RQ1.}	
Almost a quarter (24.28\%) of most popular Go projects in our dataset use \unsafe directly. 
The number of packages using \unsafe accompanies the projects growth, with an average of 17.5\% of packages using \unsafe.
In addition, project from several domains use \unsafe to some extent, but Binding projects tend to rely heavily on \unsafe. %
}

%% file: tables/rq1-descriptive.tex
 \begin{tabular}{lrr}
	\toprule
	\textbf{Statistics} &     \textbf{\#} &      \textbf{\%} \\
	\midrule
	All projects                         &  \datasetsize & 100.00 \\
	Projects that use \unsafe       &  592 &  24.28 \\
	Projects with $>$ 100 \unsafe call-sites    &    69 &   2.83 \\
	\bottomrule
\end{tabular}

%% file: RQ_UsagePatterns.tex
\textbf{Motivation.}
\rev{Thus far, we saw that almost a quarter of projects in our dataset make use of \unsafe directly in their code, which bears the follow-up question: what are developers aiming to accomplish by breaking type safety in Go? 
Knowing the most common usages of \unsafe is essential for language designers to define better safer alternatives and improve documentation.
This knowledge is also important for tool developers to focus their effort in designing solutions (e.g., static analyzers) to the most frequent cases.  
We analyze this question under two aspects:}
\begin{enumerate}
	\item \textbf{Features:} What are the most used \unsafe features?
	We want to identify the \unsafe type and functions most frequently used in practice.
	
	\item \textbf{Usage Patterns:} What developers use \unsafe for? 
	In this aspect, we focus on identifying and understanding the high-level usage patterns most commonly adopted in the most popular Go projects.	
\end{enumerate}

\subsection{What are the most used \unsafe features?}

\noindent\textbf{Approach.}
The \unsafe package is very compact, composed of one type definition (\code{Pointer}) and three exported functions (\code{SizeOf())}, \code{OffsetOf()}, \code{AlignOf()}). 
To analyse the features used by developers, we group the analysis of \unsafe usage conducted in the previous question by the four API components.

\begin{table}
	\centering
	\caption{The \unsafe operations ranked by their usage in 592 Go projects.}
	\label{tab:rq2-most-used-features}
	\include{tables/rq2-most-used-features}

\end{table}

\noindent\textbf{Results.}
As shown in~\Cref{tab:rq2-most-used-features}, the \textbf{\code{unsafe.Pointer} is the most used feature from the \unsafe API}, used by 96\% of projects that rely on \unsafe and making the bulk of 97\% of all \unsafe operation call-sites in our dataset.
The unsafe functions are used less frequently, representing together just 2.24\% of all \unsafe operations call-sites, but are still used by a considerable number of projects.
The function \code{Sizeof}, used to retrieve the size of a variable type, is used by 36\% of all projects that rely on \unsafe.

\begin{table*}[tbh]
	\caption{Usage patterns of \unsafe, manually identified in 270 project files.}
	\label{tab:rq2-usage-patterns}
	\include{tables/rq2-usage-patterns}
\end{table*}

\subsection{What are the most common \unsafe usages?}
\label{sub:rq2-unsage-usages}

\noindent
\textbf{Approach:}
Now, we proceed to understand the usage patterns of \unsafe employed by Go projects.
From the 598 projects that use \unsafe, we randomly select 270 to perform the manual analysis (see \Cref{sub:classifying-patterns}).
This number of projects provides us with a representative sample of the 598 projects that rely on unsafe with 5\% confidence interval at 95\% confidence level.
The first two authors labeled one file per project with its dominant use-case. 

Our manual analysis yielded the hierarchical label structure shown in \Cref{tab:rq2-usage-patterns}, with six major groups of usage patterns.
We aimed to be as specific as possible in our labeling, so if a \textit{Communication} use-case can be attributed to a more specific category, say \textit{System Calls}, we label the file usage as dominated by \textit{System Calls}.
In the cases where the specific category could not be identified, we label them as their respective super category.
If neither examining the code or the commit messages we encountered sufficient information on the reason of the \unsafe usage, we labeled the file as Unclear. %
We evaluate the labeling agreement with the Cohen-kappa interrater reliability~\cite{Mary:2012:kappa}. Cohen-kappa inter-rater reliability is a well-known statistical method that evaluates the inter-rater reliability agreement level for categorical scales. The result is a scale that ranges between -1.0 and 1.0, where a negative value means poorer than chance agreement, zero indicates exactly chance agreement, and a positive value indicates better than chance agreement. 
\rev{It is also important to notice that to calculate the inter-rater reliability level, we only consider the exact matches of the most specific labeling (leaf nodes in the hierarchy), but grouped the project-specific cases into a single "Other" category before the agreement calculation.}
 \rev{In our analysis, we found that both authors have substantial agreement (kappa=0.65).} Finally, the two annotators discussed the divergencies to reach a consensus.

\noindent
\textbf{Results:}
Our findings, depicted in \Cref{tab:rq2-usage-patterns}, show that \textbf{developers mostly use \unsafe for Communication to routines not written in Go (45.70\%) and to perform more efficient type casting in their programs (23.60\%)}.
Other patterns appear less common, such as inspecting the size of objects in memory (6.74\%), using \unsafe to inspect object's metadata through reflection (3.75\%), to perform atomic operations (3.37\%) and general memory manipulation (3.37\%).
Also, in 3\% of the cases, we could not pinpoint a usage pattern based on the analysis of a single file, and in 8.24\% of the cases we deemed the usage pattern too project specific to be discussed here. %
In the next paragraphs, we dive in details on each group of \unsafe pattern and explain, with examples, the likely rationale behind the breaking type-safety decisions.

\noindent
\textbf{Communication (System Call and CGO.)}
\rev{The most common usage pattern in our sample of 270 projects' files is the usage of \unsafe as a mean to communicate to routines outside of Go language, which cannot be achieved in a type-safe manner.}
The communication to systems and programs outside of Go language requires developers to specify the memory address in which such programs can read and write, to specify parameters and receive their returned objects.
For system calls, the \code{syscall} package~\cite{Go:syscall:online} offers an API to different operating systems and often requires a \uintptr with the address of a Go variable as parameters, as illustrated in Listing~\ref{lst:pattern-syscall}.

In turn, the package \cgo offers a similar set of API for developers that need to integrate with C code~\cite{Go:cgo:online}.
Similarly to the system call use-case, programs that call C code need \unsafe to write and read arbitrary memory and communicate with C code, as illustrated in Listing~\ref{lst:pattern-cgo}.
Furthermore, developers also cannot rely on the Garbage Collector from Go to release their C variables, and need to explicitly call the \code{C.free()} function to release the memory back to the system.

\rev{A possible explanation for the high frequency of patterns related to communication could be attributed to projects implementing Bindings. 
To investigate this, we computed the frequency of patterns with and without the Binding projects and noted that our results still hold: the share of communication related patterns in non-binding projects changed slightly from 57.70\% to 45.02\%. 
This shows that the use of \unsafe as a method for interacting with other systems is prevalent also in non-Binding projects.
}

\begin{lstlisting}[caption=Example of using \code{unsafe} to set the name of a process, label=lst:pattern-syscall, float, belowskip=-7pt, aboveskip=-3pt]
// Unsafe pointer as a reference to a target name
ptr := unsafe.Pointer(&name_in_bytes[0])

// Setting a process name with the pointer ptr 
_, _, errno := syscall.RawSyscall6(syscall.SYS_PRCTL, 
    syscall.PR_SET_NAME, uintptr(ptr), 0, 0, 0, 0)
\end{lstlisting}

\begin{lstlisting}[caption=Example of using \code{unsafe} to call a function in C., label=lst:pattern-cgo, float, belowskip=-7pt]
func SetIcon(iconBytes []byte) {
    // Convert to a C char type
    cstr := (*C.char)(unsafe.Pointer(&iconBytes[0]))
    // Call the function from systray.h 
    C.setIcon(cstr, (C.int)(len(iconBytes)))
}
\end{lstlisting}

\noindent
\textbf{Efficient Casting.}
In 23.60\% of the cases, developers use \unsafe as a method of bypassing compiler checks and memory copy when casting a variable to a different type.
Naturally, while there are type-safe ways of converting between valid types, developers only opt for using \unsafe to optimize runtime and memory allocation of certain functions.
The most common case of performance optimization is related to converting string to bytes and vice-versa.
Strings are immutable in Go, hence a regular type casting from a string variable requires copying the variable before the cast.
To bypass this variable copy, developers use \unsafe to change the representation of a string into a slice of bytes as illustrated in Listing~\ref{lst:pattern-optimization}. 
Since a slice of bytes can be mutated, this operation breaks the immutability of Strings as specified by the Go language, which can have far reaching consequences.

Another particular use-case for \unsafe under this category  is the more efficient marshaling functions, which accounts for 4.12\% of the use-cases. 
The standard marshaling functions are general-purpose and use reflection to identify the object to be marshalled, an operation that can be considered slow in performance critical applications.
By using \unsafe, developers are able to implement their own customized and more efficient marshaling functions.

\begin{lstlisting}[caption=Example of unsafe conversion of bytes to string., label=lst:pattern-optimization, float, belowskip=-7pt, aboveskip=-3pt]
func String2Bytes(s string) []byte {
    sh := (*reflect.StringHeader)(unsafe.Pointer(&s))
    bh := reflect.SliceHeader{
        Data: sh.Data,
        Len:  sh.Len,
        Cap:  sh.Len,
    }
    return *(*[]byte)(unsafe.Pointer(&bh))
}
\end{lstlisting}

\noindent
\textbf{Reflection.}
Reflection allows developers to inspect and modify the metadata of types at runtime, simulating some of the dynamism of dynamic typed languages.
The \code{reflect} package~\cite{Go:reflect:online} requires developers to import \unsafe to derreference an object accessed through reflection to a pointer. 
This is done by design, to enforce developers to import \unsafe when performing such unsafe operations and to prevent \code{reflect} from replicating some of the functionality of the \unsafe package.

\noindent
\textbf{Atomic Operations}
\rev{Another use-case that occurred in 3.37\% of the cases is related to the package \code{sync/atomic}~\cite{Go:atomic:online}}.
The package provides low-level atomic functions for synchronization algorithms.
As the documentation of the package poses, using \code{sync/atomic} properly requires great care from developers to be used correctly.
This package offers a way to perform atomic operations with high-performance and without any memory allocation.
Currently, there is no safe alternative to perform compare and swap operations on Go objects without memory allocation~\cite{syncatomic:proposal:online}, which explains why the package gives support to \code{unsafe.Pointer} as opposed to a type-safe alternative (e.g., \code{interface}).

\noindent
\textbf{Address Manipulation.}
In 3.37\% of the use-cases, developers use \unsafe to get the memory addresses and perform some arithmetic function. 
Most of the cases we identified are related to using the memory address as a component of a hash function, similarly to the way Java implements the object \code{hashCode()} mechanism.  

\begin{lstlisting}[caption={Code snippet of using \unsafe as a method to generate hash key, taken from the project Olric.}, label=lst:pattern-architecture-info, float, belowskip=-7pt, aboveskip=-3pt]
func (db *Olric) getHKey(name, key string) uint64 {	
    tmp := name + key	
    return db.hasher.Sum64(*(*[]byte)(unsafe.Pointer(&tmp)))	
}
\end{lstlisting}

\begin{lstlisting}[caption={Code snippet of using \unsafe to infer the system architecture, taken from the project Telegraf.}, label=lst:pattern-architecture-info, float, belowskip=-7pt]
// Verifying the size of an integer i 
if unsafe.Sizeof(i) == 4 {
    is32Bit = true
} else {
    is32Bit = false
}
\end{lstlisting}

\noindent
\textbf{Size of Object.}
In 6.74\% of the cases, the \unsafe package is used to retrieve information about the object size, a use case that is often performed to get the architecture of the underlying system, such as CPU architecture and the system endianness.
For instance, in Listing~\ref{lst:pattern-architecture-info} we present a snippet where developers verify the size of an integer \code{i} to infer whether the operating system is 32-bits or not.
\\

\conclusion{
	\textbf{Summary of RQ2:}
	The bulk of \unsafe usages are related to low-level routines that communicate to operating systems and C code (45.70\%), and to improve the performance of type casting (23.60\%).
	Less frequently, developers use \unsafe to inspect the CPU architecture (4.49\%), inspect object's metadata at runtime (3.75\%), perform atomic operations (3.37\%) and to manipulate memory addresses (3.37\%).
}

%% file: tables/rq2-most-used-features.tex
\begin{tabular}{l|rr|rr}
	\toprule
	\multirow{2}{*}{\textbf{Operations}} & \multicolumn{2}{c|}{\textbf{Projects}}  		& \multicolumn{2}{c}{\textbf{Call-sites}} \\
					  & \textbf{\#} &  \textbf{\%} 		& \textbf{\#} & \textbf{\%} \\
	\midrule
	type \code{Pointer}     &  570 & 96.28 &    177,192 &     97.85\\
	function \code{Sizeof}     &  216 & 36.49&     3,527 &      1.95 \\
	function \code{Offsetof}  &   21 &  3.55  &       312 &      0.17 \\
	function \code{Alignof}    &   11 &  1.86  &      43 &      0.02\\
	\bottomrule
\end{tabular}

%% file: tables/rq2-usage-patterns.tex
\definecolor{bluegray}{rgb}{0.4, 0.6, 0.8}

\def\scale#1{%
	\numexpr#1 / 3\relax}

\def\mybar#1{%
	{\color{bluegray}\rule{\scale{#1}mm}{7pt}}} 

\begin{tabular}{l l ll}
	\toprule
	\textbf{Usage Patterns} & \textbf{Description}	& \textbf{~~~~~~Frequency}	&			\\
	\midrule
	
	\begin{minipage}{5cm}
		\renewcommand\DTstyle{\rmfamily}
		\dirtree{%
			.2 \textbf{Communication}.
			.3 System Calls.
			.3 CGO.
			.2 \textbf{Efficient Casting}.
			.3 Type Casting.
			.3 Marshaling.
			.2 \textbf{Reflection}.
			.2 \textbf{Atomic operations}.
			.2 \textbf{Address Manipulation}.
			.3 Pointer Arithmetics.
			.2 \textbf{Size of Object}.
			.3 Getting Architecture Info.
			.2 \textbf{Unclear}.
			.2 \textbf{Others}.
		}
		
	\end{minipage}
	
	&
	
	\begin{minipage}{8cm}
		\renewcommand\DTstyle{\rmfamily} %
		\DTsetlength{0pt}{0pt}{0pt}{0pt}{0pt} %
		
		\dirtree{%
			.1 \textbf{Communicate with platforms and programs not written in Go}.
			.1 Using \unsafe to send parameters to system calls.
			.1 Using \unsafe to integrate with C code via \cgo .
			.1 \textbf{Perform more efficient type/array casting}.
			.1 Bypassing type-checking and memory copy for performance.
			.1 Using \unsafe to efficiently (de-)serialize json files.
			.1 \textbf{Access object's metadata}.
			.1 \textbf{Writing atomic operations with \code{atomic/sync}}.
			.1 \textbf{Using \unsafe to get memory addresses and copy memory}.
			.1 Perform pointer address arithmetics.
			.1 \textbf{Getting object size in memory}.
			.1 Inspecting the cpu architecture through \texttt{sizeOf(int)} method.
			.1 \textbf{Unclear usage-patterns}.
			.1 \textbf{Other use-cases (project specific)}.
		}
		
	\end{minipage}
	
	&
	
	\begin{minipage}{1.5cm}
		\renewcommand\DTstyle{\rmfamily} %
		\DTsetlength{0pt}{0pt}{0pt}{0pt}{0pt} %
		
		\begin{flushright} %
	
		\dirtree{%
			.0 \textbf{45.70\%}.
			.1 27.72\%.
			.1 17.98\%.
			.0 \textbf{23.60\%}.
			.1 19.48\%.
			.1 4.12\%.
			.1 \textbf{3.75\%}.
			.1 \textbf{3.37\%}.
			.0 \textbf{3.37\%}. %
			.1 2.25\%.
			.0 \textbf{6.74\%}. %
			.1 4.49\%.
			.0 \textbf{3.00\%}.
			.0 \textbf{8.24\%}. 
		}

		\end{flushright}

	\end{minipage}
	
	&
	
	\begin{minipage}{1.5cm}
		\renewcommand\DTstyle{\rmfamily} %
		\DTsetlength{0pt}{0pt}{0pt}{0pt}{0pt} %
		
			\dirtree{%
				.1 \mybar{45}.
				.1 \mybar{27}.
				.1 \mybar{17}.
				.1 \mybar{23}.
				.1 \mybar{19}.
				.1 \mybar{4}.
				.1 \mybar{3}.
				.1 \mybar{3}.
				.1 \mybar{3}.
				.1 \mybar{2}.
				.1 \mybar{6}. 
				.1 \mybar{4}.
				.1 \mybar{3}.
				.1 \mybar{8}. 
			}
		
	\end{minipage}
	
	\\
	
	\bottomrule
\end{tabular}

%% file: RQ_BadPractices.tex
\noindent\textbf{Motivation:}
\rev{In the previous RQs, we found that \unsafe is commonly used in Go projects and there are six main reasons that motivate developers to break type safety. 
While using \unsafe is dangerous by definition, we want to investigate the potential risks that using this package entails to software projects in terms of software issues. 
These potential risks are relevant for Go developers, \revminor{so that they can identify the types of issues their project are likely to face when breaking type safety}.
Documenting the risks is also useful for tool developers and researchers, who can focus on strategies of mitigation, identification, and repair of these issues. 
Therefore, we investigate the risks of using \unsafe under the following two aspects:}
\begin{enumerate}
	\item \textbf{Invalid usage:} We analyze the \unsafe call-sites with static analyzers to identify invalid pointer conversions in the projects' source code.
	Invalid pointer conversion is a well-known pitfall that projects depend on \unsafe are at the risk of falling into.
	
	\item \textbf{Unsafe-related issues:} We extract and classify issues related to \unsafe from the projects repositories.
	This analysis will give us in-depth insights of real problems faced by projects that use \unsafe.
	
\end{enumerate}

\vspace{-0.1in}
\subsection{How common is invalid pointer conversion in projects that use \unsafe?}
\label{sub:rq3-invalid-pointer-conversion}

\noindent
\textbf{Approach:}
We run the tool Go Vet on all projects that use \unsafe. We were able to automatically analyze 221 projects, due to problems during the project build.
Several projects use \unsafe to integrate with external systems (C code and Operating Systems), hence, these projects in our dataset need external dependencies which cannot be automatically resolved with our building process. Once we run the tool Go Vet, we count the number of reported cases of invalid pointer.

\noindent
\textbf{Results:}
Regarding invalid usage of \unsafe, initially the Go Vet tool reported invalid pointer conversions in 16 out of 221 analyzed projects (\Cref{tab:rq3-reports}). 
\revminor{Upon close inspection, we found that cases found in two projects were false-positives, as the conversion to \uintptr and back to pointer occurred in the same expression, as permitted by the official \unsafe documentation~\cite{Go:unsafe:online}}.
Therefore, \textbf{14 out of 221 projects (6.6\%) had clear invalid pointer conversions}, with functions receiving a \uintptr as parameter and converting it back to pointer inside the function or performing unsafe pointer arithmetics.
Albeit occurring in a minority of the investigated projects, this analysis shows that some popular projects still fall into the most well-documented pitfalls of using \unsafe.
\revminor{We note that these results should be interpreted as a lower bound of invalid point conversions in these 211 projects. 
That is, because Go Ver uses a set of patterns to identify invalid pointer conversion, the tool will likely miss cases that do not fall into the known patterns. 
As a consequence, there could be more than 14 cases in the 221 analyzed projects, but there is no practical way to assess how many of such cases were not retrieved by the tool (false negatives).}

\begin{table}
	\centering
	\caption{Number of projects with invalid pointer conversions. We did not observe projects with a mix of false-positive and true-positive invalid pointer conversions.}
	\label{tab:rq3-reports}
	\include{tables/rq3-reports-local}
\end{table}

\subsection{What kind of unsafe-related issues projects that use \unsafe have?}
\label{sub:rq3-issues}

\noindent
\textbf{Approach:}
After filtering issues through the keyword search in the title \revminor{("unsafe", "un-safe")}, we identify 286 issues from 119 projects.
Our manual inspection revealed that only 103 unsafe-related issues from 63 projects were in fact, related to the \unsafe package.
We conduct our manual analysis on this set.  
Again, we use Cohen-kappa inter-rater \cite{Mary:2012:kappa} to evaluate the labeling agreement between the two annotators. We found that both annotators had a moderate agreement (Cohen-kappa=0.55) after the first labelling round.
Later, both annotators discussed all divergences and reached a consensus in the classification.

\begin{table*}[tbh]
	\caption{Issues related to \unsafe identified in 63 projects. We present both the number of issues (\#) and the number of projects in which they occur (\# proj.).}
	\label{tab:rq3-unsafe-related-issues}
	\include{tables/rq3-unsafe-related-issues}

\end{table*}

Our manual classification of the unsafe-related issues yields a scheme shown in \Cref{tab:rq3-unsafe-related-issues}. 
We group the issues into 10 categories comprised of Bug Fixing and Project Maintenance issues.
We consider as Bug Fixing the issues that were open due to runtime errors or bugs found in the project code base.
Project maintenance, on the other hand, are issues created with the goal of improving the project, by adding new features or refactoring the code to improve the overall quality and reduce maintenance costs. 
In the next paragraphs we discuss, with examples, each category of unsafe-related issues identified in our dataset of the 63 projects.
\\

\noindent
\textbf{Unsafe Restriction.}
The most common unsafe-related issue in our dataset, found in 20 projects, is related to external environmental restriction of the use of the \unsafe package. 
In most cases, this is related to the Google App Engine, a platform for cloud development that restricts the usage of \unsafe for any Go code running in the platform~\cite{AppEngin9:online}, due to safety reasons.
For example, a developer in the project GJSON wrote: \snippet{I wanted to use this package within a Google App Engine project, and due to package "unsafe" being used, it is not compatible}~\cite{Removedu37:online}.
In many of these cases, the solution found for project maintainers was to provide a version of their package without \unsafe dependency, or to remove the \unsafe dependency in favor of a safer alternative.
\revminor{However, we also found cases where the usage of \unsafe is so widespread in the project's code, the maintainers were not willing to remove it.}
In such cases, there is an encouragement that users fork the project repository to create a new safe-version of their packages, e.g., this developer wrote: \snippet{The atomic swap is used all over the place during transactions, so I don't think we'd want to take a change that removes it, but you could make a fork and replace unsafe with a mutex basically to make it less performant but safe}~\cite{googleap87:online}.

\noindent
\textbf{Runtime Errors.}
In 16 projects, we encounter issues that were created due to runtime errors caused by the misuse of \unsafe package.
The most common type of runtime error, found in 6 projects, is related to crashing errors due to bad pointers, e.g. \snippet{Prometheus crashes and hangs on `fatal error: found bad pointer in Go heap}~\cite{Promethe40:online}.
Such errors can be caused by the Garbage Collector releasing unintended variables, mismanagement of operations that read and write memory and possibly other causes.
Another cause of crashing bugs in some projects were related to the conversion between different type layouts. 
As a developer points out in an issue \snippet{I suspect the problem is that there is no guarantee that the alignment is correct after the pointer conversion}~\cite{Flakyuns29:online}.
The misuse of \unsafe has also been reported to cause data corruption,  \snippet{Unsafe use of unsafe that leads to data corruption}~\cite{Usafeuse2:online} due to breaking string immutability in the string to bytes conversion. 
Furthermore, we also found reports indicating that the program did not crash during execution but the wrong usage of \unsafe has led the program to produce wrong results: \snippet{The combination of this version of siphash and use of unsafe.Pointer to obtain a byte slices caused back-to-back ast\#Term.Hash calls to return different values!}~\cite{Removeus35:online}.
These errors are difficult to diagnose and replicate: 5 of such issues mentioned problems to reproduce the runtime error due to the non-deterministic nature of the problem.
As a developer describes: \snippet{But someday, when using my JS package, I stumbled upon an unexpected behavior in one Lua function\footnotemark it is not even an error, just that the string.gsub function isn't behaving correctly}~\cite{strangee68:online}. 
All such reports corroborate with the expected risks of breaking \revminor{type safety}, which may cause crashing errors, data corruption, wrong behavior and in many cases are difficult to diagnose and replicate.

\footnotetext{The Lua function was implemented in Go, through gopher-lua.}

\noindent
\textbf{Wrong Usage.}
We group in this category, 14 bugs found on 12 projects caught by maintainers or collaborators while inspecting the code, with no report of runtime issues.
The most common case, found in 5 projects, is related to slice conversion:
\snippet{While not likely to occur in the wild, changes to how GC inlines functions in 1.12 creates the possibility for data loss when serialization\_littleendian.go uses unsafe to change the type of slices}~\cite{Keephead73:online}.
Slice conversions also suffer from invalid pointer conversion issues, having the Garbage Collector release the slice in the midst of the conversion. 
A commonly employed solution for this problem is to explicitly inform the runtime system to not release the slice header during the conversion, by calling the function \code{runtime.KeepAlive}.
However, \code{runtime.KeepAlive} is not always a valid solution~\cite{Go101:2020:KeepAlive} and can still yield memory corruption in some particular cases~\cite{GoPlay:online}. %
We also found issues related to traditional invalid pointer conversion, as described by a developer in an issue: \snippet{The following code is invalid because it puts a non-pointer value into a pointer-type, if the GC finds this non-pointer it will crash the program}~\cite{Invalidu72:online}.
Furthermore, a particular issue related to \unsafe usage draws our attention.
The issue was reported as a security related in the project \code{nuclio}, where  convertion from string to bytes raised the possibility of exposing parameters from an internal library, as a developer reports: \snippet{Without going through their code this [issue] is likely due to buffer reuse, which is a race condition at the least and a security issue at the worst}~\cite{security31:online}.

\noindent
\textbf{Static/Dynamic Check Violation.}
In total, 7 issues from 6 projects were opened due to static and dynamic checks identifying suspicious usage of \unsafe. 
In 5 out of 7 cases, the issue was raised by the newest Go 1.14 version (released in February, 2020) which incorporated in the compiler a more robust invalid pointer conversion checker, which instruments the code to find violations to the \unsafe rules dynamically. 
For example, a developer reported:
\snippet{Go 1.14 introduces new runtime pointer checking, enabled by default in race mode which verify that the rules for using unsafe are followed}~\cite{runtimee32:online}.
This check is performed if the program is compiled with a race detector (flag \code{-race}) in Go 1.14 and emits a fatal error if a violation is found, forcing developers to fix the problem. 
The other two cases were motivated by \code{go vet}, the static analyzer we used to identify invalid pointer conversions in our previous analysis.

\noindent
\textbf{Breaking Changes and Portability Issues.}
While less frequent, we also found issues related to breaking changes (3 projects), and related to portability issues (2 project).
\revminor{For instance, in 2011 a change in the function \code{sizeOf} of the package \unsafe created several bugs in a project.}
As a developer describes \snippet{In tip, unsafe.Sizeof has been changed to return uintptr. This causes a variety of issues during the build of walk}~\cite{unsafeSi28:online}. 
We also found portability issues related to projects that use \unsafe when used in different Operating Systems.
For instance, in one issue a developer reported that \snippet{unsafe (marshaler and unmarshaler) test fails on big endian architectures}~\cite{KnownIss30:online}, indicating unsafe implementation worked on powerpc (32-bit) but did not work on s390x (64-bit) architectures.
These issues corroborate with the \unsafe package documentation, which clearly states that portability and compatibility guarantees \revminor{do not hold} when using \unsafe.

\noindent
\textbf{Project Maintenance.}
We report Project Maintenance issues related to \unsafe grouped in four groups. 
Such issues are not really bugs found in projects and do not represent a direct risk for the projects. 
Instead, they represent tasks that are open by collaborators to improve project maintenabililty and are reported here in our study for completeness.
The most common issue, found in 12 projects, were created to suggest the removal of \unsafe to mitigate the  risks of using \unsafe, e.g., \snippet{Our benchmarks also show that there is no significant difference between safe and unsafe. This allows to remove optimizations with unsafe and simply rely on plain code generation}.~\cite{Nounsafe3:online}.
Moreover, we found 2 projects with issues related to isolating the \unsafe dependency to a reduced number of packages in a project.
\rev{This indicates that some developers are keen to reduce the reliance on \unsafe due to its inherent risks.} 
\revminor{While there is an effort to keep \unsafe usage at minimal levels in some projects, we also found that developers expand their API to support the \code{unsafe.Pointer} type ("Support for channels, maps and unsafe.Pointers"~\cite{Supportf3:online}).}
As shown in RQ2, developers frequently rely on \unsafe for optimizations and we found issues open in 4 projects proposing a \revminor{refactoring} to more efficient code that uses \unsafe.

\noindent
\textbf{Fix-patterns to unsafe related issues.} \revminor{We investigated the recurrence of fix-patterns that could be used to fix unsafe related issues automatically. We focused in identifying fix patterns related to issues of Wrong Usage and Runtime Errors, as these are closer to the traditional functional bugs. From the 32 issues, we found the majority of 24 issues to be entirely context-dependent. These fixes were related directly to the business logic of the program and are hard to generalize. The remaining 8 issues were fixed following fix-patterns related to unsafe type-casting, which we discussed in \Cref{sec:background}: developers should perform operations using a single expression to avoid having the GC release variables in use.  }

All the examined bug issues are exclusively caused by the use and misuse of \unsafe.
The main takeaway from this analysis is that projects that do not break \revminor{type safety} and depend on type-safe third-party packages are free of encountering the issues we discussed in this analysis, such as crashing errors caused by bad pointers, have portability and compatibility issues or have their code restricted to deploy in different environments.

\conclusion{
	\textbf{Summary of RQ3:}
	We found suspicious cases of invalid pointer conversions in 14 out of 211 projects. More importantly, projects that use \unsafe face several unsafe-related issues, from having their deployment restricted (20 projects), experiencing runtime errors that are hard to reproduce (16 projects), introducing bugs due to misuse of \unsafe (12 projects). 
}

%% file: tables/rq3-reports-local.tex
\begin{tabular}{lrr}
	\toprule
	\textbf{Statistics}							   &	\textbf{\#}	& 	\textbf{\%}	\\
	\midrule
	All projects analyzed			&	221		&	100.0\%		\\
	Projects with invalid 
	pointer conversions 			& 14		  & 6.3\%			\\
	Projects with reported false-positives 					& 	2			& 			0.9\%		\\
	\bottomrule
\end{tabular}

%% file: tables/rq3-unsafe-related-issues.tex
\begin{tabularx}{\linewidth}{llXrr}
	\toprule
	\textbf{Task} &\textbf{Issue category} &  \textbf{Reason behind the Issue}  &\textbf{ \#}  & \textbf{\# Projects}           \\
	\midrule
	\multirow{6}{*}{Bug Fixing} 
	& Unsafe restriction  & Deployment environment restricts the use of \unsafe &  30 &         20 \\
	& Runtime errors & Wrong usage of \unsafe caused runtime and crashing errors & 18  &         16 \\
	& Wrong usage & Bug found in the code related to wrong usafe of \unsafe (preemptive) & 14 &         12 \\
	& Static/Dynamic Check Violations & Bug in the code found by static code analyzers (Go Vet or Go1.14) &   7 &          6 \\
	& Breaking Changes & Issue due to breaking changes introduced in \unsafe  & 4 &          3 \\
	& Portability Issues & Program did not work in different architectures & 3 &          2 \\

	\midrule
	\multirow{4}{*}{Maintenance} 
	& Remove unsafe & Replace \unsafe with a safer implementation variant & 13 &         12 \\
	& Extending \unsafe support & Add support to unsafe.Pointer & 6 &          6 \\
	& Using \unsafe for optimization & Optimize code using \unsafe  & 6 &          4 \\
	& Isolate unsafe usage & Move \unsafe code to a dedicated package &  2 &          2 \\

	\bottomrule
\end{tabularx}

%% file: Discussion.tex
\rev{
In this section, we delve into the design decisions behind the exposition of \unsafe on standard Go packages~(\Cref{sub:design-decision}) and discuss the performance benefits of common unsafe optimizations~(\Cref{sub:performance-optimization}). Then, based on our results we discuss a set of recommendations to improve the \revminor{safety} of the language (\Cref{sub:implications}) and describe an exchange we had with the Go team about our study (\Cref{sub:go-team}).
}

\vspace{-0.1in}
\subsection{Design decisions of APIs exposing \unsafe}
\label{sub:design-decision}
\input{Discussion_officialAPIs}

\vspace{-0.1in}
\subsection{Performance benefits of \unsafe casts}
\label{sub:performance-optimization}

\input{Discussion_optimization}

\vspace{-0.1in}
\subsection{Recommendations}
\label{sub:implications}
\input{Implications}

\vspace{-0.1in}
\subsection{Perspectives of the Go Team}
\label{sub:go-team}
\input{GoTeam_feedback}

%% file: Discussion_officialAPIs.tex
The majority (56\%) of the \unsafe usage patterns are linked to packages from standard libraries (\code{syscall}, \code{cgo}, \code{reflect} and \code{atomic}).
Given the risks associated with breaking \revminor{type safety}, we wanted to investigate why standard Go packages exposed \unsafe in their APIs.
Hence, we went through the list of standard packages in Go\footnote{List of standard packages in Go:~\url{https://golang.org/pkg/}}, and listed the packages that expose \unsafe by either using \texttt{unsafe.Pointer} or \uintptr in their API.
The standard library of Go has a total of 147 packages, covering a variety of different functionalities, from the language core to marshaling of different data formats. 
 We found that \textbf{7 out of 147 packages expose \unsafe in their APIs (see \Cref{tab:discussion_apiexposeunsafe})}.
We observe that these seven packages expose \unsafe for the following reasons:

\begin{enumerate}
	
	\item[R1] \rev{The package provides access to low-level implementation which needs to be handled carefully by developers. In this case, having the requirement of importing \unsafe forces developers to understand the unsafeness of the package itself.
	This is the case of the \code{reflect} package, which could return an unsafe pointer in the function \code{Pointer()}, but instead returns an \uintptr to force developers to import \unsafe to do the conversion~\cite{Go:reflect:online}.}

	\item[R2] \rev{The package provides low-level pointer operations that cannot be efficiently implemented with safe pointers.
	This is the case of the package \code{atomic} that performs atomic operations in primitives and pointers. 
	It is not possible to dereferrence or alter the address of a safe pointer in Go, hence the package needs to expose \unsafe to swap pointers in a fast and atomic way~\cite{syncatomic:proposal:online}.}
	
	\item[R3] \rev{The package gives access to non-portable or system-specific implementation, such as \code{syscall}~\cite{Go:syscall:online} and \code{cgo}~\cite{Go:fix:online}. In both cases, the reliance on syscalls and C code already puts the program into non-compatibility territory, as operating systems and external dependencies might change their APIs without further notice.}
\end{enumerate}

\begin{table}
	\centering
	\caption{\revminor{Packages from Go standard library that have at least one function that requires the import of \unsafe. We describe the reasons R1, R2, and R3 in the text of Section 7.1.}}
	\label{tab:discussion_apiexposeunsafe}

\input{tables/discussion_apisexposingunsafe}
\end{table}

\noindent \rev{Overall, we find that only a small share of standard Go packages requires the use (or import) of the package \unsafe. 
The exposition of \unsafe from such packages seems to be driven by a design decision to warn developers of the risk of certain operations as well as to provide the flexibility and performance that cannot be achieved in a type-safe manner with the current Go language.}

%% file: tables/discussion_apisexposingunsafe.tex
\begin{tabularx}{\linewidth}{l X l}
	\toprule
	\textbf{Package} 	& \textbf{Description} & \textbf{\revminor{Reason}}\\
	\midrule
	\code{os}				&	Provides a platform-independent interface to operating system functionality.				& \revminor{R1, R3}									\\
	\code{reflect}			&	Implements run-time reflection, allowing a program to manipulate objects with arbitrary  types.												& \revminor{R1}	\\
	\code{runtime}		&	Contains operations that interact with Go's runtime system.							& \revminor{R1}							\\
	\code{cgo}			&	Runtime support for code generated by the cgo tool.											& \revminor{R3}			\\
	\code{debug}	&	Contains facilities for debugging programs.								& \revminor{R1}						\\
	\code{atomic}	&	Provides low-level primitives for implementing synchronization algorithms.		& \revminor{R2}												\\
	\code{syscall}	&	Contains an interface to the low-level operating system primitives.			& \revminor{R3}												\\
	
	\bottomrule
\end{tabularx}

%% file: Discussion_optimization.tex
\begin{table}
	\centering
	\caption{Example of pull-requests open to improve the performance of functions by using \unsafe. }
	\label{tab:performance-improvement-casts}
	\input{tables/performance-improvement-casts}
\end{table}

\rev{
Developers use \unsafe as a method for implementing faster functions in 23.6\% of the usages clasified in RQ2.
We want to investigate level of performance gain that can be obtained by breaking type safety in Go. 
We looked into the six issues classified as ``Using \unsafe for optimization" in \Cref{sub:rq3-issues} and identified that in four cases, the developers reported the optimization obtained by using \unsafe.
We report these in \Cref{tab:performance-improvement-casts}. 
The improvement observed varies from no improvement in some functions to up to 90\% execution time and 50\% less memory allocation.
While these number seem impressive, they report the time spent and memory allocated on the function optimized alone (micro-benchmarks). 
We found no end-to-end performance tests to evaluate how much of an impact an improvement would have on a software in production.}

\begin{figure}
	\centering
	\begin{subfigure}[t]{0.48\linewidth}
		\centering
		\includegraphics[width=\linewidth]{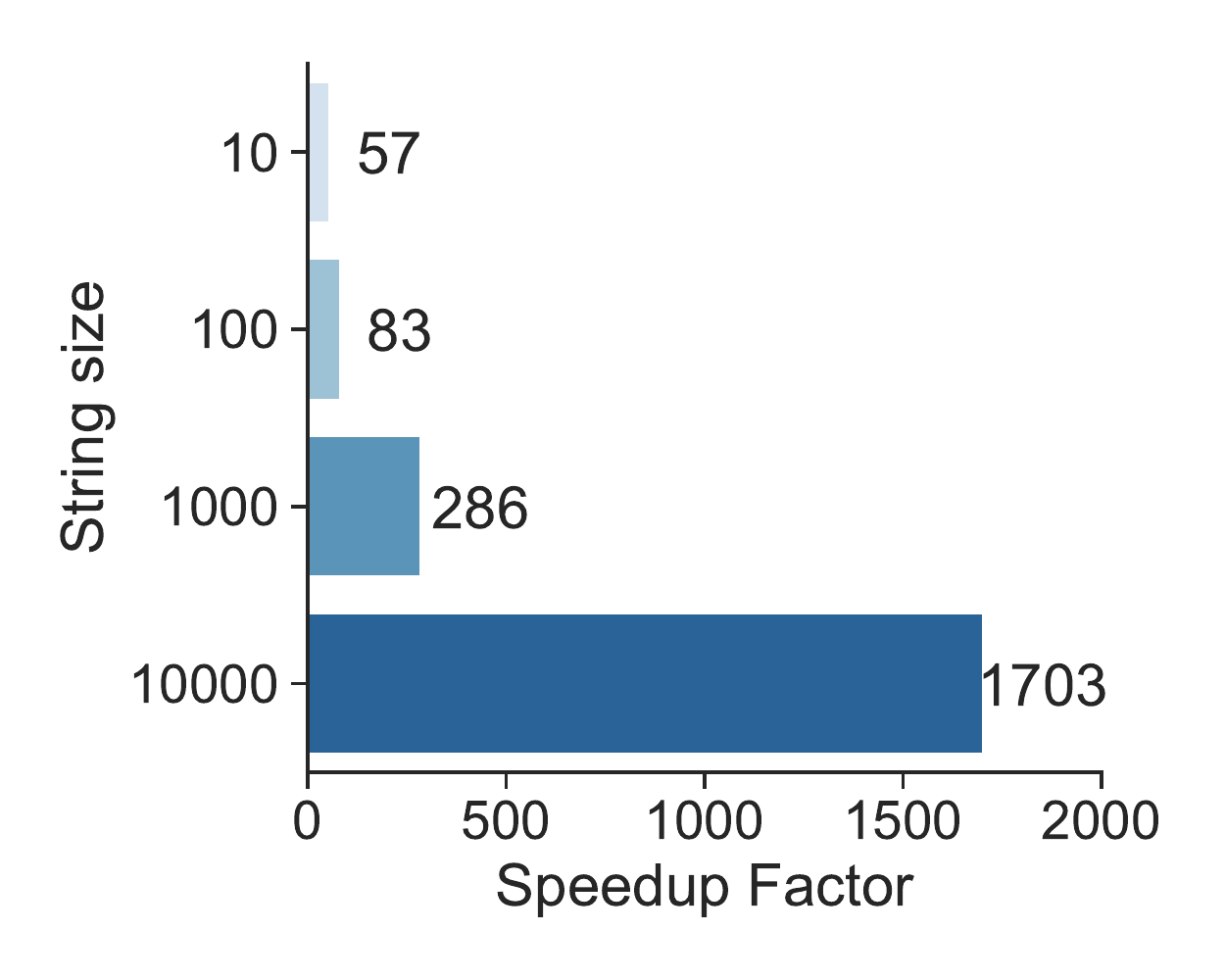}
		\caption{String to bytes.}
		\label{fig:string2bytes}
	\end{subfigure}
	~
	\begin{subfigure}[t]{0.48\linewidth}
		\centering
		\includegraphics[width=\linewidth]{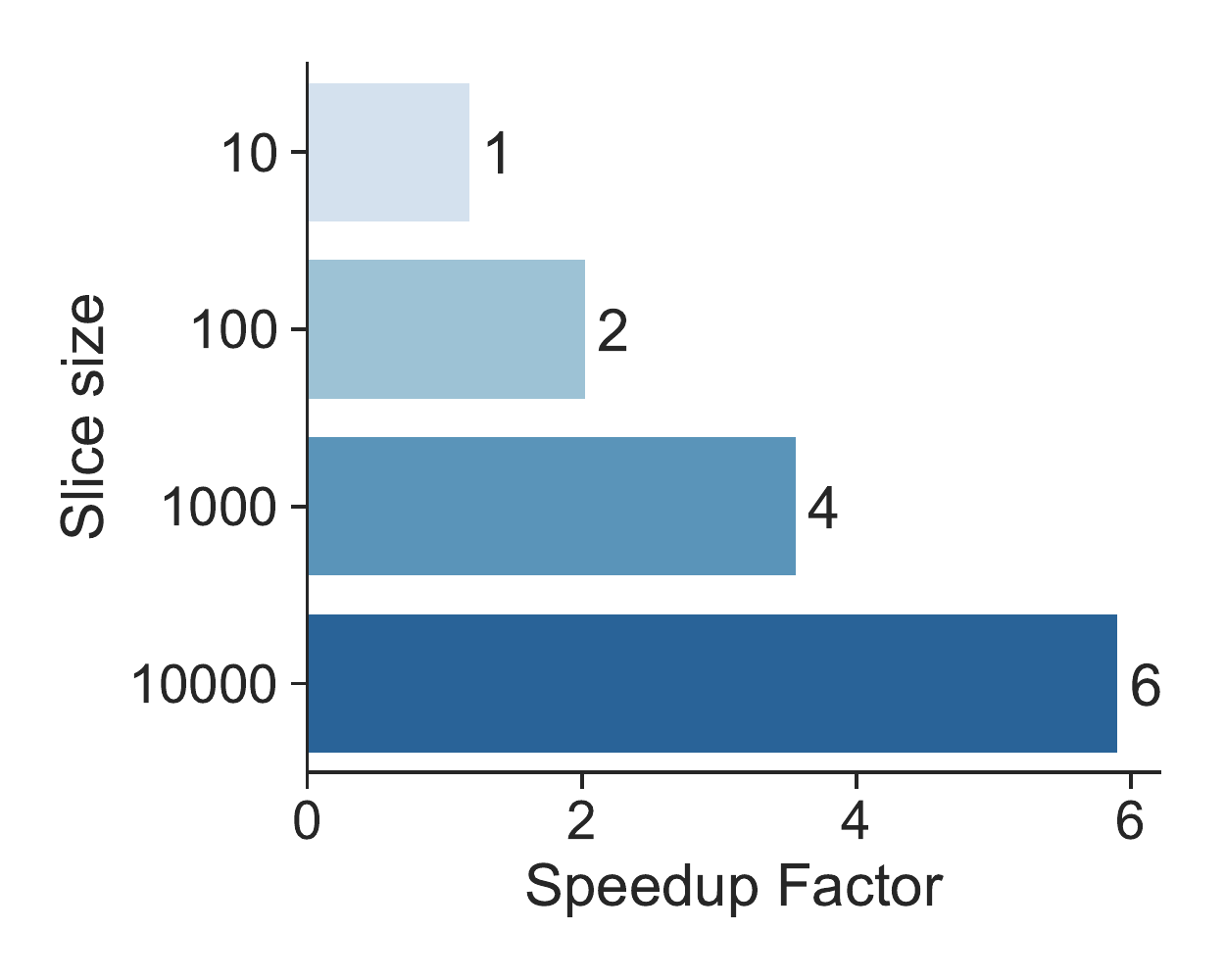}
		\caption{Slice of bytes to integers.}
		\label{fig:slicecasting}
	\end{subfigure}
	
	\caption{Performance speedup factor obtained by using \unsafe casting compared to type-safe alternatives in a 2.3 GHz Quad-Core Intel i5 machine. 
		In both cases, the time and memory allocation complexity of the type-unsafe alternative is O(1) compared to the O(n) of the safe implementation.}
	\label{fig:discussion-performance}
		\vspace{-0.15in}
\end{figure}

As we discussed in \Cref{sub:rq_usagepatterns}, the most common code unsafe function used to optimize code is the conversion of strings to bytes and vice-versa.
This is a well-known method for optimizing performance but has harmful \revminor{potential} consequences of breaking string immutability specified by the Go language.
For instance, trying to modify the bytes after the conversion from a string will cause a segmentation violation error.
In fact, most code snippets come with a recurring comment of ``Use at your own risk" indicating that developers are aware of the risks but choose to use this as an optimization method.
\revminor{To illustrate the performance benefits of this optimization method, we crafted two micro-benchmarks to compare the performance of 1) safe vs unsafe string conversion and 2) safe vs unsafe slice conversion. }
We run the benchmark using Go native harness benchmarking package~\cite{Go:testing:online} to reliaby measure the steady-state performance of these functions under different loads.  
We show in \Cref{fig:string2bytes} the speedup obtained in a micro-benchmark with varying string sizes when converting string to bytes.
Even at smaller string sizes (10 characters), the speedup obtained through bypassing the string copy reaches a factor of 57x, which may explain why so many popular Go projects contain this function in their code-base.

Developers also use \unsafe to speed-up slice casting operations. 
As discussed in the Wrong Usage issue in \Cref{sub:rq3-issues}, this method also has inherent complexities. 
If not done correctly the conversion can cause the program to crash, particularly if the types converted do not have the same memory layout. 
We also craft micro-benchmarks to evaluate the performance speedup of using the \unsafe cast and found that, unlike string to bytes conversions, the performance benefits are not clear-cut (see \Cref{fig:slicecasting}).
Unsafe slice conversions are up to 6x faster on a slice with 10 thousand integer elements, which may not justify the risks of the type-breaking conversion in this particular case.

%% file: tables/performance-improvement-casts.tex
\begin{tabular}{l l r}
	\toprule
	\textbf{Optimization}	& \textbf{Reported performance improvement}	& \textbf{Ref.}		\\
	\midrule

	Slice casting			& 	60\% to 90\% faster execution time		& 	\cite{Optimization1:online}		\\
	String to Bytes			& 0\% to 40\% less memory allocation		& \cite{Optimization2:online}\\
	String to Bytes			& 50\% less memory allocation					& \cite{Optimization3:online} \\			
	String to Bytes			& 25\% faster exec time + 27\% less allocations			& \cite{Optimization4:online} \\

	\bottomrule
\end{tabular}

%% file: Implications.tex
Our results show that \unsafe usage is widespread, motivated by low-level software integration and performance optimization. However, it puts projects at the risk of several issues.
Next, we draw a series of recommendations from our results that could help make Go language more safe.

\noindent
\textbf{Inclusion of more powerful static analyzers.}
Currently, the tool Go Vet focus on identifying invalid pointer conversions, one of the main issues with using \unsafe.
However, researchers and practitioners can develop analyzers that attempt to find other invalid cases of \unsafe usage, such as using \code{unsafe.Pointer} to \revminor{dereference} a nil pointer or access a memory address beyond the allocated memory space.
For instance, statically identifying null pointers is a problem well investigated in languages like C and C++~\cite{Hovemeyer:2005:NullPointer, eigler2003mudflap}, and similar approaches could be employed in Go to identify unsafe nil pointer conversion.
In fact, the newest Go release (1.14) already provides a more robust set of checks, by embedding a compile option that instruments the code to capture violation of safety rules.

Furthermore, our study shows that the majority of \unsafe usages are concentrated on a handful of patterns. 
The catalog of usages provided by our study can be employed in static analysis tools to identify misuses of widely-used patterns. 
For instance, two widely-used patterns, the slice and bytes-string conversions should draw special attention and become target \revminor{patterns for static analyzers}. 
Developers often implement such conversions without guarding for invalid pointer conversion, causing their programs to crash and behave erroneously.

\noindent
\textbf{Improve documentation on frequent \unsafe usages.} 
\rev{Our investigation showed that the conversion of string to bytes (and vice-versa) and slice types are a very common optimization method (present in at least 19\% of projects as shown in RQ2). } %
However, the official documentation of \unsafe only briefly touches the fundamentals of the issues developers encounter when writing this code.
This is further corroborated by discussions in mailing lists related to Go on how to properly convert string to bytes \cite{Clarific27:online} and proposals to clarify the usage of \unsafe when performing syscalls~\cite{proposal36:online}.
While Go language maintainers have expressed that including such conversions in the standard API is not desired~\cite{Featurep4:online}, due to the issues of breaking string immutability, an official statement on how to properly convert between string and bytes could help developers at writing correct code conversion and finding the ill-implemented variants.

\noindent
\textbf{Inclusion of Generics.}
\revminor{We found some usages of \unsafe to be primarily motivated by the type flexibility that a support for generics would provide, e.g., Marshaling which is present in at least 4.12\% of the projects.}
Without generics, generic functions (e.g., customized ways of sorting a slice) are accomplished with interfaces, reflection or code generation~\cite{GoIsthee45:online}.
The standard marshaling package relies on runtime reflection to encode and decode json to Go structs, which can negatively impact the performance of object serialization.
The inclusion of generics would allow for more efficient encoders/decoders without relying on \unsafe, as the 
compiler could generate code automatically based on type specified by developers in their generic functions~\cite{Go2Gener96:online}.

\noindent
\textbf{\revminor{Planning breaking changes to \unsafe in the Go language.}}
There exists several proposals that if implemented in Go 2.0 may impact programs that currently use \unsafe~\cite{proposal:unsafe1:online,proposal:unsafe2:online}, introducing breaking changes in the language.
Our study shows that a quarter of the most popular Go projects use \unsafe in their code, and given their notoriety, it is expected that such projects are used by a large share of the Go community.
We build a dependency graph using the Go List command, considering only the project in our dataset (\datasetsize most popular Go projects), and found that 40\% of the projects either use \unsafe directly or depend directly on a project that uses \unsafe.

Fortunately, our study also shows that the majority of the \unsafe usage in popular Go projects is well-localized in code: most projects concentrate their \unsafe in at most 2 files.
This indicates that the cost for updating the \unsafe usages to comply to possible breaking changes in Go 2.0 version should be manageable for most projects.
Still, language designers could mitigate the cost for adopting the new language version by communicating the adopted proposals in advance to the community, or even better, by employing a tool for migrating some of the \unsafe usages to the newest package version within the Go Fix command~\cite{Go:fix:online}.
The Go Fix is a tool created with the sole purpose of migrating old APIs to new ones in the case of breaking changes in the language, and can be used to update \unsafe usages from valid Go 1.x code to Go 2.x.
\rev{Our methodology can also be employed on a larger set of Go projects to assess the impact of the breaking change in the entire Go ecosystem and better aid language designers at planning mitigation measures.}

%% file: GoTeam_feedback.tex
\newcommand{\teamquote}[1]{\textit{``#1"}}

The end-goal of our work lies in making the Go language safer and easier to use by developers. 
Naturally, this goal is way beyond what this study can achieve on its own and requires that both the language designers and developers are aware of the risks of using unsafe and adopt measures to improve its usage in the long run.	
Coincidentally, once we made a pre-print of this work available online, we were reached by a member of the Go Team, Matthew Dempsky, with positive remarks on our study.
Matthew is a maintainer of the Go compiler and responsible for the implementation of \texttt{checkptr}, the dynamic checker deployed with Go 1.14 reported in \Cref{sub:rq3-issues}.
With his help, we circulated the pre-print version of this paper to his colleagues to receive feedback on any result from the paper the Go team deemed to be surprising or useful. 
We summarize the highlights of this exchange in the following:

 \begin{enumerate}
	\item \rev{The widespread usage of unsafe was received as a surprise by the Go team. The usage of \unsafe by 24\% of the projects is significantly higher than the Go team expected. As Matthew stated, \teamquote{Other team members were more optimistic that developers would avoid or could implement their project without using package unsafe. I think this result will help dispel that misconception and justify spending more time on making package unsafe easier to use.}}

	\item \rev{The usage patterns and some of the mistakes were known by the Go team and the results were consistent with what the members expected. Still, their relative frequencies provided new insights about their importance. \teamquote{I think knowing the frequencies will be useful in prioritizing improvements to unsafe.Pointer.}}

	\item \rev{The Go team has also pointed out that the trends observed in the historical analysis were novel and insightful, showing that projects tend to keep up their \unsafe dependencies during their entire development.}

 \end{enumerate}

\noindent\rev{Overall, the interest and the feedback we received from the Go Team highlight a positive outcome of our study. In particular, the predominance of \unsafe usage in popular Go projects (RQ1) showed that breaking \revminor{type safety} is common and developers need better support to mitigate its risks.
}

%% file: Related_work.tex
There exists a plethora of work that investigate how developers use certain features of the programming language. 
In this section, we discuss the work that is most related to our study. We grouped prior work into work related to unsafe features in programming language and studies on mining repositories to assess language features.

\vspace{-0.1in}
\subsection{Unsafe Language Features}

\revminor{Some prior studies investigates how developers use low-level code features in different programming languages.}
Nagappan~\others~\cite{Nagappan:2015:GoTo} investigate how developers use the \code{goto} statements in a representative sample of C programming files.
Motivated by the harmful stigma of \code{goto} statements, the authors qualitatively investigate how such command is used in modern C code. The investigation showed that developers nowadays limit themselves to use \code{goto} only in very specific circumstances, such as error handling and cleaning up resources, and that it does not appear to be harmful in practice.

Some other studies focus on investigating the usage of dynamic features on programming languages.
Mastrangelo~\others~\cite{Mastrangelo:2019:Casting} investigated how and when developers use casting on thousands of Java projects.
The results show that casting is widely-used by Java developers and that half the casts are not guarded locally to ensure against runtime errors. 
Similarly to our study, this investigation also cataloged a common set of use-case patterns that can help language maintainers and tool developers to better accommodate the most common usages of dynamic casting.
Callau~\others~\cite{Callau_DynamicTypingSmallTalk} investigated how dynamic and reflective features are employed in SmallTalk, by surveying a thousand SmallTalk projects.
While SmallTalk is a dynamically-typed language, the authors reported that dynamic features are not widely used by SmallTalk developers. 
\rev{In fact, the two most pervasively used features are the ones other static languages implement, indicating a more conservative approach with seemingly unsafe features than the ones we observed in popular Go projects.}

\rev{One of the works most closely related to ours is the investigation of unsafe APIs in Java by Mastrangelo~\others~\cite{Mastrangelo:2015:JavaUnsafe}.
In their work, the authors investigated how developers use the \code{sun.misc.Unsafe}, a class that exposes low-level and unsafe features, violating Java safety guarantees.
Differently from the \unsafe package in Go, the \code{sun.misc.Unsafe} API gives access to more than type safety violations, for instance, developers can violate method contracts by throwing a checked exception undeclared by a method.
Their investigation found that only a small share (1\%) of software artifacts use the unsafe API in their code directly.
In contrast, we found that 24\% of the most popular Go projects rely on unsafe to some extent, showing that using unsafe APIs seems to be more common in Go than in Java.}

\rev{Another closely related work investigated the use of Unsafe Rust~\cite{Evans:Rust:2020} in a large set of Rust libraries.
Unsafe Rust is in fact a language embedded in Rust, but it gives access to similar functionalities as the \unsafe package, such as C-style pointer operations and access to arbitrary memory locations, which are not guaranteed by the compiler to be safe.
The authors have identified that 30\% of the Rust libraries use the keyword unsafe and more than half of them depend on another library that is unsafe. 
This results are in line with the findings we observed in our study. 
Our study contributes to this body of work, that investigate the practices of breaking \revminor{type safety}~\cite{Mastrangelo:2015:JavaUnsafe, Evans:2020:Rust} in programming languages and provides insights and recommendations to make programming languages more safe.}

\vspace{-0.1in}

\subsection{Mining Repositories of Usage of Language Features}

Several studies empirically investigate how developers use different language features through mining software repositories. 
Mazinanian~\others~\cite{Mazinanian:17:Lambdas} mined software repositories to investigate how Java developers use Lambda in their programs. 
Similarly, Costa~\others~\cite{Costa:17:Collections} profiled how developers select and tune their data structures in Java programs, showing that developers only rarely tune their data structures and prefer the standard implementations, despite having better performance variants available.
Krikava~\others~\cite{Krikava:2019:ScalaImplicits} investigate the usage of implicits - a language feature that allows developers to reduce boilerplate code - in Scala programs.
The results showed a pervasive use of the feature in Scala projects hosted in GitHub. Guilherme and Weiyi~\cite{GuilhermeI:CPC2017} performed an empirical study to examine the prevalence of exception-handling anti-patterns in Java and C\# projects. Their findings showed that all studied projects contain exception handling anti-patterns in their source code. Wang~\others~\cite{WangSANER2019} studied the evolution regular expressions over time from the different aspects. Their results showed that the use of regular expressions is stable in the development history of the studied GitHub projects.

The aforementioned work mines software repositories to quantitatively assess the level of adoption of language features from developers. 
Our study shares part of this mining repository methodology, such as selecting studied projects, parsing the code to get quantitative results and manually analyze the usage patterns for a qualitative assessment.

%% file: Threats_to_validity.tex
This section describes the threats to the validities of our
study.

\vspace{-0.1in}
\subsection{Internal validity}

Threats to internal validity are related to experimenter bias and errors.
To identify \unsafe usages, we build a customized parser which could miss or introduce false positives in our analysis. To mitigate this threat, 
we first took special care to identify all possible ways \unsafe can be used in Go code, by inspecting import statements for \unsafe in a number of projects. \revminor{Second, during our manual examination of the detected cases of \unsafe to catalog the usage patterns, we only found a single case (1/270) of false positive, which give us a the confidence in the accuracy of our customized parser.}
Developers had defined a custom object named ``unsafe" and called a method from the object, filtered in by our parser. 
Third, during our experiments, we did not observe any errors when conducting both the analysis on the last snapshot nor the historical analysis, which could indicate we could miss real \unsafe usages (false negatives). 
\revminor{Also, our parser is compatible with all Go version since Go 1.0.0. 
Therefore, we believe the results yielded by our unsafe parser will hold.}
We also use the tool Go Vet to identify suspicious cases of invalid pointer conversions. While we have manually assessed that 2 out of 16 cases were false positives, the tool could miss real cases due to the complexity of statically analyzing the code.
Hence, our results with Go Vet should be interpreted as a lower bound of invalid pointer conversions.

Furthermore, we conduct two major manual analyses in our study, to investigate the most used usage patterns and to classify unsafe-related issues. 
\rev{In the classification of the \unsafe usage patterns, we include a single random file per project, hence, our analysis is based on the usage of a single randomly selected file from a representative sample of projects.
This ensured we performed a cross-project analysis and did not bias our usage pattern catalog towards a minority of projects that use \unsafe extensively (100+ files).  
To investigate the representativeness of our analysis, we fully classified all \unsafe usages of 30 randomly selected projects of our dataset, accounting for 168 files that depend on \unsafe.
\revminor{This sample analysis showed that 66\% of the projects contain only one \unsafe usage pattern, and 86\% of the projects contain fewer than two usage patterns in their code-base.}
Furthermore, we observed a similar ranking as the one shown in \Cref{sub:rq2-unsage-usages} when analyzing the distribution of usages under a different perspective: the ranking of \unsafe usages based on the percentage of projects with at least one case of the usage pattern.
Upon the results of our validation, we believe the ranking reported in RQ2 holds even under other (reasonable) sampling strategies.}

Regarding the analysis of unsafe-related issues, it is important to note that the categories of issues obtained in this analysis are not exhaustive or necessarily representative. 
We only investigate issues that have the keyword "unsafe" in its title and hence are bound to miss many unsafe-related issues.
Hence, our analysis is a lower bound of possible unsafe-related issues, providing qualitative support to the risks of using \unsafe entails to software projects.

\vspace{-0.1in}
\subsection{External validity}
Threats to external validity are related to the generalizability of our findings.
Our investigation focused on the most popular Go projects. 
Go has established itself as a programming language for high-performance infrastructure projects and the projects in our dataset reflect that (e.g., there is a relative low number of front-end projects). 
The \unsafe package is used for low-level implementation and optimization, hence its prevalence is expected to decrease on less popular projects or on a more diverse set that may not have the pressure for high-performance code. 
We argue that, while not fully generalizable to all Go software projects, our dataset contains the most influential projects of the current Go landscape. Also, in our analysis, we examine open-source Go projects that are hosted on GitHub. Thus, our results may not generalize to proprietary projects.

%% file: Conclusion.tex
In this paper, we present the first study on the usage of \unsafe in Go programs. 
We conduct a mix-method analysis of the prevalence of \unsafe in popular open-source Go projects, investigate what developers aim to achieve by breaking \revminor{type safety}, and evaluate some of the real risks projects that use \unsafe are subjected to.
Our results have shown that the use of \unsafe is prevalent, with one in four Go projects breaking type safety primarily for communicating with programs outside of Go language (C code and Operating Systems), and optimizing type casting functions. 
Developers have made the conscious effort of keeping \unsafe restricted to a selected set of packages in their project, but report a series of unsafe-related issues due to misusage of \unsafe.
Their projects may have the deployment restricted by shared cloud environments such as the Google App Engine, developers report crashing errors and wrong results due to invalid pointer conversion, and in some cases the program produces the wrong behavior due to non-deterministic causes.

Our study can be used as empirical evidence on the state of usage of \unsafe in Go and help motivate further tools and language developments that could make Go safer. 
We suggest that special attention should be given towards creating tools that identify further bad practices related with the package, similarly to how Go 1.14 introduces more robust checks for invalid pointer conversions, as developers seem to struggle in writing valid unsafe code.
Language maintainers can also mitigate the encountered issues by documenting official unsafe snippets to be used by the community, in particular on type casting of string and bytes, widely used as a \revminor{means} of optimization.
Furthermore, the level of adoption of \unsafe should also be taken into consideration when planning future versions of the language, as the package as risky as it is, seems to be integral to the Go community.